\PassOptionsToPackage{comma,numbers,sort&compress,super}{natbib}

\documentclass[10pt,aps,prl,twocolumn,showpacs,amsmath,amssymb,floatfix,superscriptaddress]{revtex4-1}
\setcitestyle{super}
\usepackage[version=4]{mhchem}
\usepackage[T1]{fontenc}
\usepackage{multirow}
\usepackage{graphicx}
\usepackage[dvipsnames,table]{xcolor}
\usepackage{booktabs}
\usepackage{amsmath}
\usepackage{xr}
\usepackage{blindtext}
\usepackage{gensymb}
\usepackage{silence}
\usepackage{tabularray}
\UseTblrLibrary{booktabs}

\color{darkgray}

\begin{document}

\title{Twin-boundary-induced nonrelativistic spin splitting}

\author{Kristoffer Eggestad}
\affiliation{Department of Materials Science and Engineering, NTNU - Norwegian University of Science and Technology, NO-7491 Trondheim, Norway}

\author{Marc Vila}
\affiliation{Materials Sciences Division, Lawrence Berkeley National Lab, 1 Cyclotron Road, Berkeley, CA 94720, USA}
\affiliation{Molecular Foundry, Lawrence Berkeley National Lab, 1 Cyclotron Road, Berkeley, CA 94720, USA}

\author{Sverre M. Selbach}
\email{selbach@ntnu.no}
\affiliation{Department of Materials Science and Engineering, NTNU - Norwegian University of Science and Technology, NO-7491 Trondheim, Norway}

\author{Sinéad M. Griffin}
\email{sgriffinh@lbl.gov}
\affiliation{Materials Sciences Division, Lawrence Berkeley National Lab, 1 Cyclotron Road, Berkeley, CA 94720, USA}
\affiliation{Molecular Foundry, Lawrence Berkeley National Lab, 1 Cyclotron Road, Berkeley, CA 94720, USA}

\sloppy

\begin{abstract}

Nonrelativistic spin splitting (NRSS) in compensated magnetic materials is drawing considerable attention due to its potential impact in next-generation spintronic devices. While NRSS is typically restricted to materials with particular symmetry constraints, here we demonstrate, using density functional theory (DFT) and tight-binding transport calculations, that twin boundaries can induce NRSS in magnetic systems where it is otherwise forbidden. We focus on two representative material systems: the tetragonal perovskite oxide \ce{BiCoO3} with $90\degree$ ferroelastic domain walls, and the rhombohedral layered delafossite-type oxide \ce{CoO2}, supporting $71\degree$, $109\degree$, and $135\degree$ twin boundaries. Our results reveal that, if these boundaries coexist with ferromagnetic domain walls, they consistently produce NRSS similar to that of d-wave altermagnets, with nodal surfaces dictated by the underlying symmetry of the supercell containing the twin boundary. Tight-binding models further elucidate how the NRSS and derived transport properties scale with domain size and density. Our results put forward twin boundary engineering as a versatile route to realize and control spin splitting in a broader class of materials.
\end{abstract}

\maketitle

Magnetic materials with compensated magnetic order and a spin-split electronic structure offer a promising route toward next-generation spintronic devices \cite{Jungwirth2016, Baltz2018, Manchon2019}. They combine the ultrafast magnon dynamics characteristic of antiferromagnets \cite{Hortensius2021, Zelezny2017} with the ability to generate and manipulate spin currents efficiently \cite{Shao2021, Dong2022, Qin2023, Chen2023, Samanta2025}. A particularly intriguing manifestation of this physics is nonrelativistic spin splitting (NRSS), which has recently attracted considerable attention as a means to achieve large spin splittings whose origin lies in exchange interactions and crystal-field effects rather than spin-orbit coupling \cite{C5CP07806G, Naka2019,doi:10.1126/sciadv.aaz8809,PhysRevB.99.184432, PhysRevB.102.014422, doi:10.7566/JPSJ.88.123702, Yuan2021}. 

In compensated magnets, NRSS arises when the two magnetic sublattices are not related by inversion ($\mathcal{P}$) or translation ($\tau$) symmetries combined with time-reversal ($\mathcal{T}$). Among the strategies to realize this condition, \textit{altermagnets}~\cite{Smejkal2022A} have emerged as a broad class of materials exhibiting NRSS, with numerous examples already experimentally verified \cite{Krempasky2024, Zhu2024, Lee2024, Osumi2024, Reimers2024, Yang2024, Zeng2024CrSb, Dale2024, Jiang2025, Zhang2025}. In these systems, the magnetic sublattices are connected by a composite symmetry of crystal rotation or mirror ($\mathcal{M}$) and time reversal. This combined operation dictates the momentum dependence of the spin splitting and defines nodal planes where spin degeneracy is preserved. Consequently, recent efforts have concentrated on identifying and designing bulk crystals whose space-group symmetries permit such altermagnetic order \cite{Sodequist2024, Wan2025, Bandyopadhyay_et_al:2025}. 

NRSS, however, need not be restricted to bulk symmetry. Previous studies have shown that surfaces and interfaces can modify magnetic order and introduce additional symmetry breaking \cite{PhysRevX.14.021033, PhysRevLett.134.146703, 7brd_lynv, Ho2025}. While these works focused on magnetic reconstruction at material surfaces, here we extend the concept of NRSS itself beyond the bulk, showing that its underlying symmetry requirements can be fulfilled through the deliberate design of structural motifs beyond the unit cell. 

Here, we establish a general symmetry-based framework for realizing NRSS in such engineered motifs. Starting from the requirement that the two magnetic sublattices must not be related by $\mathcal{PT}$ or $\mathcal{\tau T}$ symmetry, but may instead be connected through a rotation-$\mathcal{T}$ or mirror-$\mathcal{T}$ operation, we identify \textit{twin boundaries} as a natural geometry satisfying these conditions. When twin boundaries coincide with ferromagnetic domain walls, the boundary enforces a rotation symmetry equivalent to that responsible for NRSS in bulk altermagnets, thus generating the same momentum-dependent spin texture without requiring an intrinsically altermagnetic crystal structure. 

To demonstrate this principle, we perform density functional theory (DFT) calculations for the ferroelastic perovskite \ce{BiCoO3} and the layered delafossite \ce{CoO2}. Both exhibit twin-boundary-induced NRSS despite bulk symmetries that forbid altermagnetism. Complementary tight-binding quantum-transport simulations show that the spin current associated with this domain-wall-induced NRSS is robust against variations in domain size but diminishes with increasing wall density due to interfacial scattering. These results broaden the range of materials expected to host NRSS and highlight twin boundaries as an experimentally accessible and versatile platform for engineering spin-split electronic states.

\begin{figure}[t]
\centering
\includegraphics[width=0.9\linewidth]{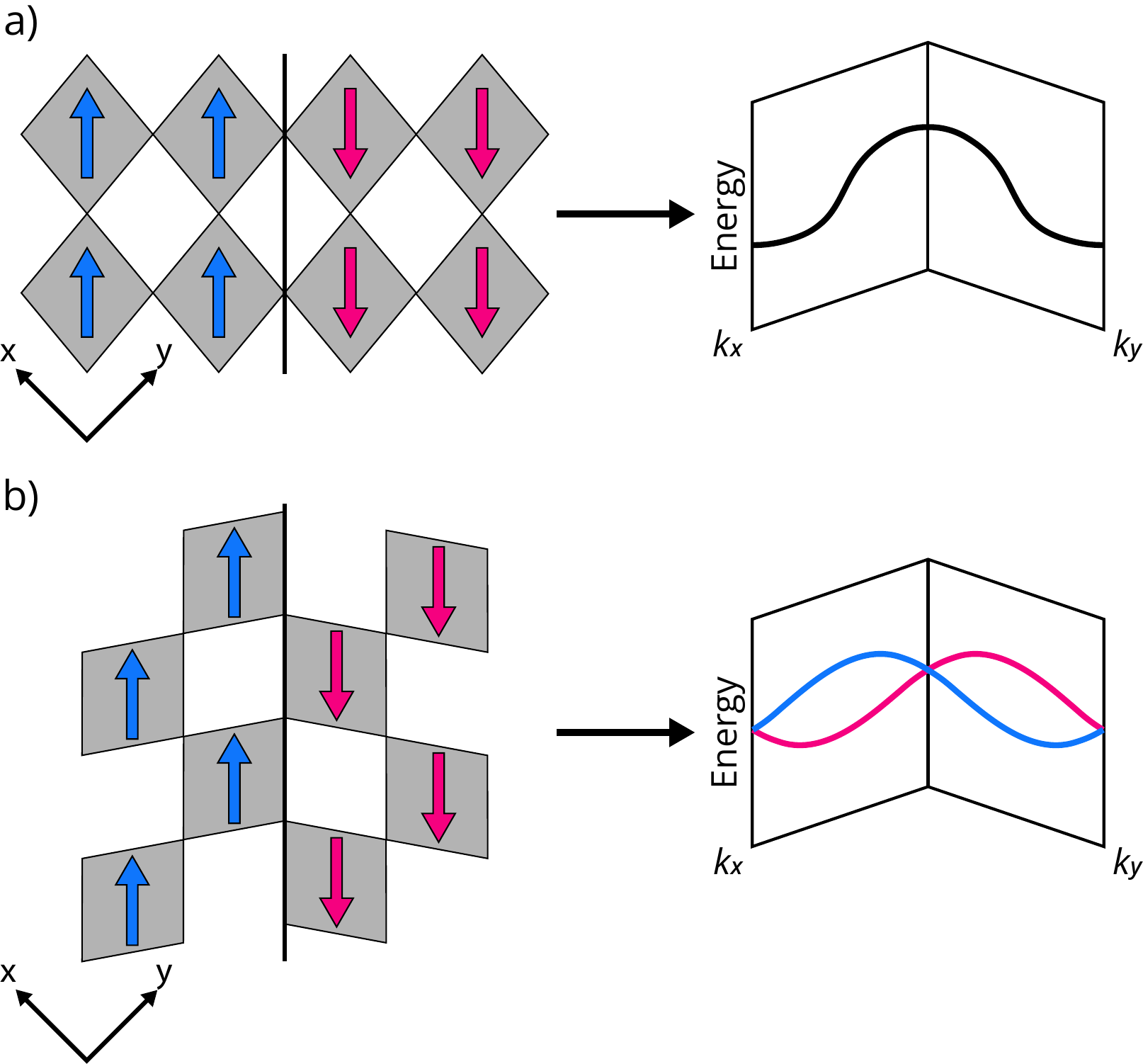}
\caption{Illustration of nonrelativistic spin splitting induced by twin boundaries. (a) A ferromagnetic domain wall alone does not produce spin splitting. (b) When a ferromagnetic domain wall coincides with a twin boundary, spin splitting can occur.}
\label{fig:F1}
\end{figure}

The central symmetry principle of our work is illustrated in Fig.~\ref{fig:F1}, which compares two types of ferromagnetic domain walls separating regions of opposite magnetization. In Fig.~\ref{fig:F1}\,(a), the two magnetic domains share the same crystallographic orientation; the wall is purely magnetic, and the domains remain related by either $\mathcal{PT}$ or $\mathcal{\tau T}$ symmetry, resulting in spin-degenerate bands. In Fig.~\ref{fig:F1}\,(b), the domains are rotated relative to one another, so that the magnetic wall coincides with a structural twin boundary. Translation and inversion no longer compensate the magnetic sublattices, enabling NRSS to appear. The twin boundary and the plane perpendicular to it act as (glide) mirror operations connecting the two domains and defining the location of spin-degenerate nodal planes. Because twin boundaries can be atomically sharp~\cite{Salje01031999, wruck_thickness_1994}, these symmetry operations remain well defined in real materials, offering a practical route to engineer NRSS through coupled magnetic-structural domain architectures.

We next validate this symmetry-based framework using first-principles calculations on two representative systems. The first is the multiferroic perovskite \ce{BiCoO3}, where twin boundaries can form as ferroelastic domain walls. The second is layered \ce{CoO2}, in which twin boundaries have been experimentally observed~\cite{JIANG2020105364,exp_twin}. In both cases, the domain-wall geometry satisfies the symmetry conditions derived above, allowing a direct demonstration of our design principle. Computational details are provided in the Supporting Information (SI).

\begin{figure}[t]
\centering
\includegraphics[width=0.95\linewidth]{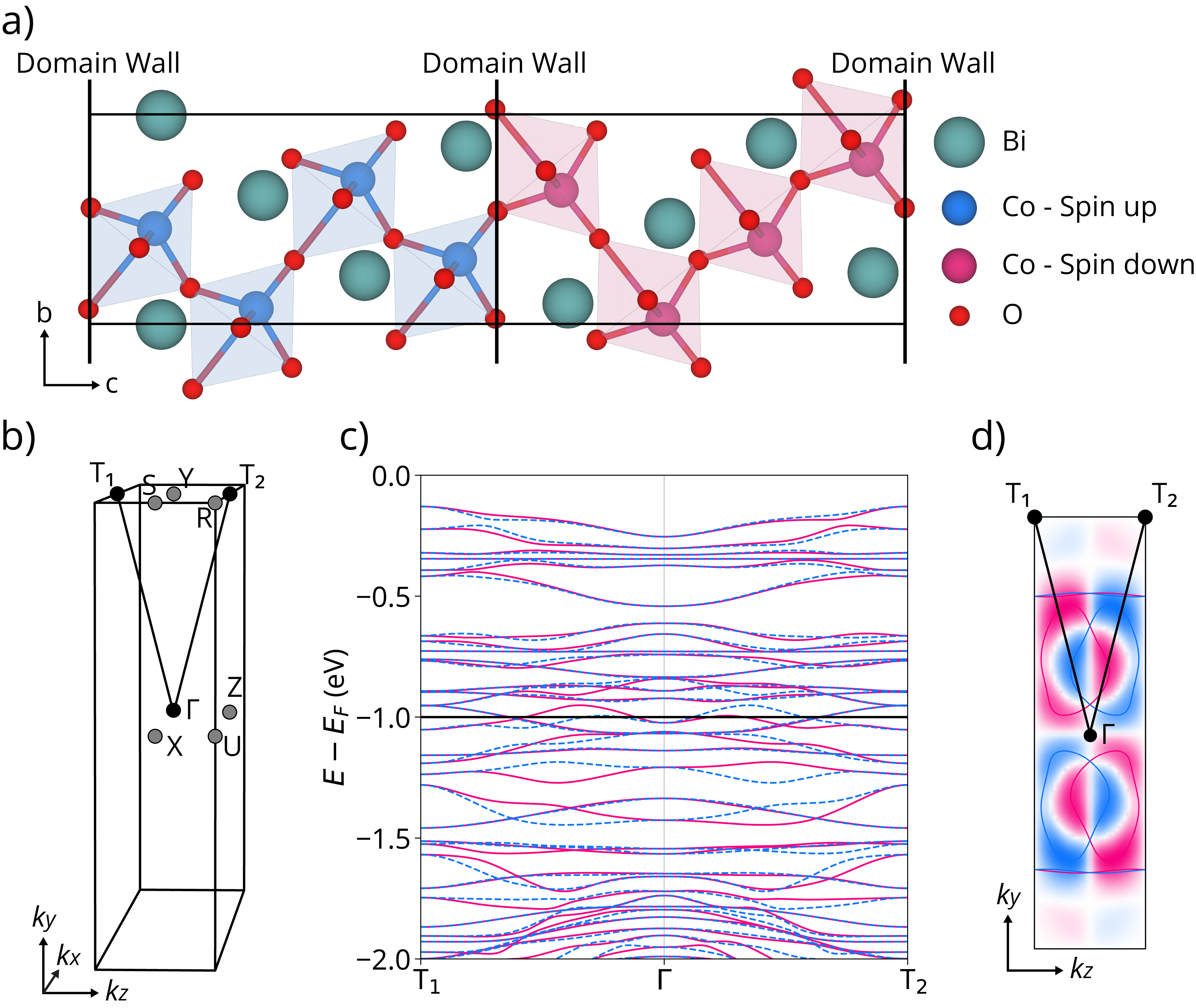}
\caption{(a) Crystal structure of a supercell containing eight formula units of \ce{BiCoO3}, displaying ferromagnetic, ferroelastic, and ferroelectric domain walls, indicated by thick black vertical lines. Due to periodic boundary conditions, two of each domain wall type are present in the cell. (b) Corresponding Brillouin zone, where the high-symmetry path (T$_1 \rightarrow \Gamma \rightarrow \text{T}_2$) used in the band structure is indicated. (c)  Calculated electronic band structure along the Brillouin zone path shown in (b). The solid (pink) and dashed (blue) lines depict the two spin polarizations. The Fermi level is set to 0~eV with the solid black line marking the energy at $-1.0$ eV below the Fermi level.  (d) Constant energy cut in the $k_y-k_z$ plane at $-1.0$ eV below the Fermi level. The solid colored lines represent intersections of spin-up and spin-down bands with the energy cut (black line in (c)), and the contour plot shows the broadened bands as calculated using Eq.~1 in SI.}
\label{fig:F2}
\end{figure}

\ce{BiCoO3} is a multiferroic material that crystallizes in the tetragonal $P4mm$ space group at ambient conditions~\cite{doi:10.1021/cm052334z}. It exhibits a spontaneous electric polarization and a C-type antiferromagnetic order, with spins aligned ferromagnetically along the $c$ axis and high-spin Co$^{3+}$ ions carrying large local moments~\cite{PhysRevMaterials.2.104417,doi:10.1021/cm052334z}. The spin-up and spin-down sublattices are related by $\mathcal{\tau T}$ symmetry, precluding NRSS in the bulk. Importantly, \ce{BiCoO3} is isostructural to \ce{PbTiO3}, a well-known ferroelastic ferroelectric that forms $90^{\circ}$ domain walls~\cite{Catalan2011,10.1063/1.360561}. Motivated by this analogy, we construct a $90^{\circ}$ ferroelastic domain wall in \ce{BiCoO3}, shown in Fig.~\ref{fig:F2}\,(a), containing four magnetic sites per domain. As a proof of concept, we impose ferromagnetic domains, while adopting the lowest-energy head-to-tail ferroelectric configuration (see SI Tab.~S2). The resulting symmetry of the supercell transforms the sublattice-exchanging operations into combinations of glide mirrors and time reversal, thereby inducing NRSS. Specifically, the mirror parallel to the wall enforces the $\mathcal{M}_c\mathcal{P}\tau_{c,1/2}\mathcal{T}$ operation, and the mirror perpendicular to it imposes $\mathcal{M}_b\tau_{c,1/2}\mathcal{T}$, where the mirror subscript denotes its normal axis and the translation subscript specifies the fractional translation direction. Because the supercell belongs to the polar space group $Pmc2_1$, symmetry predicts NRSS with two nodal planes intersecting at $\Gamma$, analogous to $d$-wave altermagnetism in bulk crystals. The calculated band structure (Fig.~\ref{fig:F2}\,(c)) indeed shows momentum-dependent NRSS with a sign reversal between $k$ and $-k$ and a maximum splitting of $\sim$0.09 eV. The constant-energy contour in Fig.~\ref{fig:F2}\,(d) reveals spin-degenerate nodal planes at $k_y=0$ and $k_z=0$, consistent with our expectations from symmetry.  

\begin{figure*}[ht]
\centering
\includegraphics[width = 1.0\linewidth]{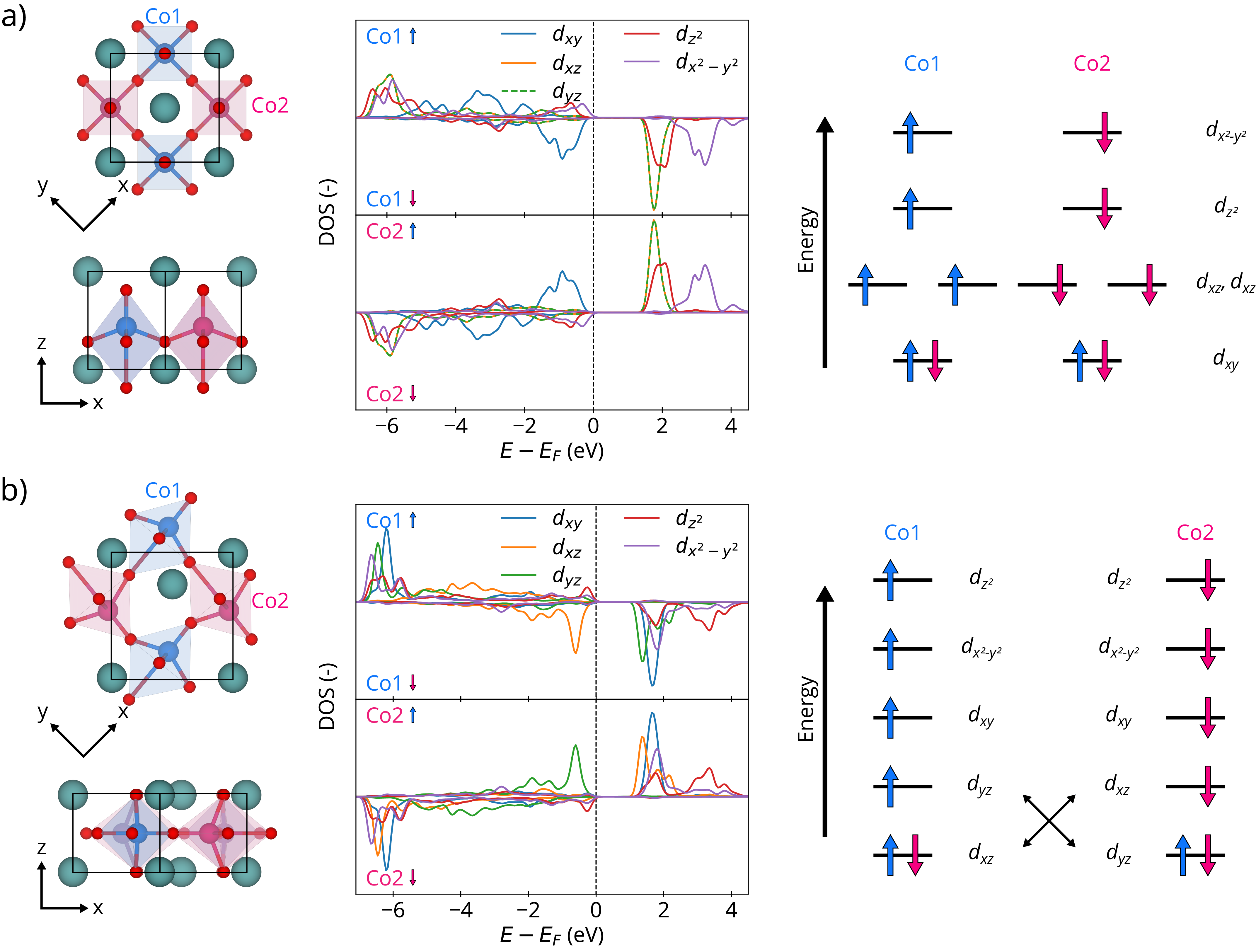}
\caption{Calculated dlectronic density of states (DOS) for the d-orbitals of the two Co atoms in \ce{BiCoO3}: (a) in the primitive unit cell with C-type antiferromagnetic order, and (b) in the minimal engineered cell containing one formula unit per ferroelastic and ferromagnetic domain. The corresponding structures are shown to the left of each DOS plot, along with the axis used for orbital projection. To the right, schematic diagrams illustrate the d-orbital filling.}
\label{fig:F3}
\end{figure*}

Although symmetry analysis alone can predict the presence of NRSS, understanding its microscopic origin requires considering both the spin and orbital order. In many altermagnets, NRSS emerges from compensated magnetic sublattices with antiparallel spins that occupy orbitals of distinct character~\cite{Smejkal2022A,Vila2025}. Such orbital differentiation can result from crystal-field distortions or electronic correlations~\cite{Vila2025,Leeb2024}. In \ce{BiCoO3}, the crystal-field variations across the twin boundary naturally create this orbital asymmetry.

Fig.~\ref{fig:F3}\,(a) shows the bulk ground-state structure of \ce{BiCoO3}, where Co$^{3+}$ ions adopt a high-spin $d^6$ configuration. The $d_{xy}$ orbital is fully occupied, while the remaining $d$ orbitals are half-filled. Because the two Co sites in the primitive cell experience identical crystal fields, their orbital-projected densities of states are equivalent apart from spin orientation, resulting in no momentum-dependent NRSS.

In contrast, the $d$-orbital-resolved densities of states for a $90^{\circ}$ head-to-tail domain wall (Fig.~\ref{fig:F3}\,(b)) reveal clear orbital polarization. The $x$ and $y$ axes are rotated by $45^{\circ}$ relative to the wall, with $z$ perpendicular to the interface. The tetragonal distortion of the CoO$_6$ octahedra aligns along either the $x$ or $y$ direction, generating distinct local crystal fields on the two Co sites. As a result, Co1 preferentially occupies the $d_{xz}$ orbital, whereas Co2 favors $d_{yz}$, as depicted in the crystal-field diagram of Fig.~\ref{fig:F3}\,(b)~\cite{Vila2025}. This alternating orbital occupation, coupled to the antiparallel spin alignment, produces the NRSS observed in Fig.~\ref{fig:F2}\,(c). Layer-resolved densities of states (SI Fig.~S6) confirm that this orbital ordering persists for larger domain sizes.

Finally, we note that the head-to-tail ferroelectric configuration introduces a small additional asymmetry at the domain wall due to the local polar distortion. This subtle structural breaking slightly differentiates the Co magnetic moments on the head and tail sides, producing a tiny net magnetic moment ($\sim 10$ m$\mu_B$) per wall, which cancels across opposite domain walls. While this effect does not alter the overall $d$-wave-like NRSS character, it offers a potential handle for controlling local magnetic moments through electric fields and subsequent ferroelectric polarization switching (see SI Fig.~S12).

Having shown in \ce{BiCoO3} that twin boundaries provide the necessary orbital symmetry differentiation between magnetic sublattices to produce NRSS, we next turn to a system where this mechanism can be more systematically tuned. The delafossite-type oxide \ce{CoO2}, structurally related to the battery material \ce{Li_{x}CoO2}~\cite{doi:10.1126/science.1122152,https://doi.org/10.1002/adma.201000717}, provides an fitting material platform for this purpose. Delafossites of the general formula \ce{ABO2} accommodate a wide range of A- and B-site cations~\cite{doi:10.1021/ic50098a011}, leading to diverse electronic and magnetic behavior. Their structure consists of edge-sharing \ce{BO6} octahedral layers separated by A-site cations or, in \ce{CoO2}, by vacant sites, resulting in quasi-two-dimensional magnetism that is highly sensitive to lattice distortions and interlayer coupling. Depending on the structural details, delafossite compounds exhibit frustrated antiferromagnetism, ferromagnetic planes coupled antiferromagnetically along the stacking direction, or even diamagnetism~\cite{doi:10.1021/acs.jpcc.0c07847,C4CP03052D}.

\begin{figure}[bt]
\centering
\hspace{-10pt}
\includegraphics[width=0.95\linewidth]{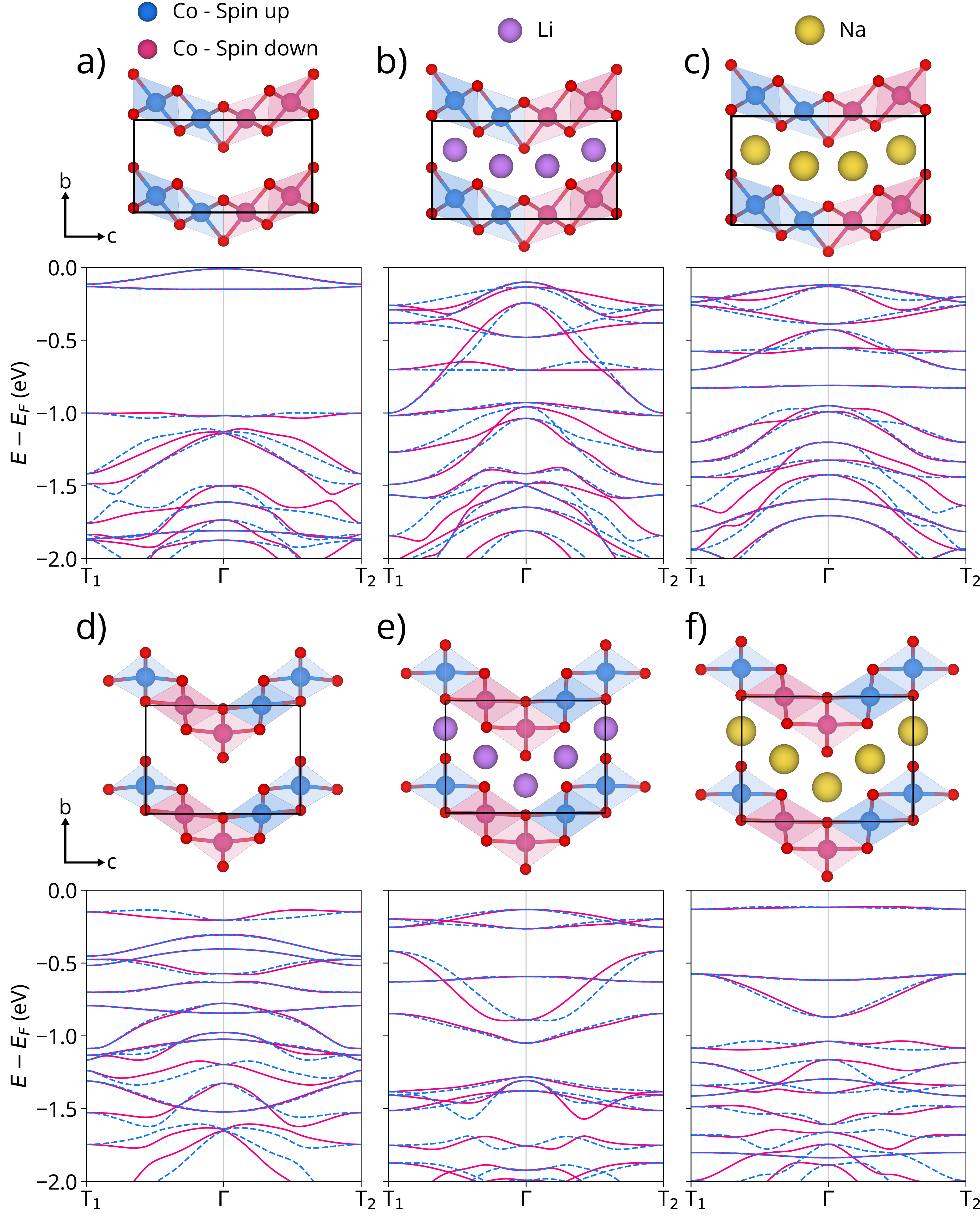}
\caption{Crystal structures and calculated spin-polarized electronic bands of \ce{CoO2} (a, d), \ce{LiCoO2} (b, e) and \ce{NaCoO2} (c, f) for the $135\degree$ (a-d) and $109\degree$ (d-f) twin boundary. The Fermi level is set to 0~eV in all figures.}
\label{fig:F4}
\end{figure}

Stoichiometric \ce{CoO2} crystallizes in the rhombohedral $R\bar{3}m$ structure and has been experimentally shown to host twin boundaries~\cite{WOS:000304628500009,JIANG2020105364}, making it a natural testbed for domain-wall-induced NRSS. By combining domains of different orientations, three distinct twin boundaries can form, approximately $135^{\circ}$, $109^{\circ}$, and $71^{\circ}$, as illustrated in Figs.~\ref{fig:F4}\,(a), \ref{fig:F4}\,(d), and SI Fig.~S10\,(a). Among these, we calculate the $109^{\circ}$ boundary to be the most energetically stable (see SI Tab.~S2) consistent with experiment, whereas the $135^{\circ}$ variant offers a simpler structural motif for analysis. Each twin boundary contains a glide plane perpendicular to the interface and a mirror plane parallel to it. When opposite spin orientations are imposed in adjacent domains, the resulting supercells adopt $Pmmn$ or $Pmma$ symmetry, both of which satisfy the conditions for NRSS analogous to $d$-wave altermagnetism~\cite{Smejkal2022A}. The calculated band structures (Figs.~\ref{fig:F4}\,(a,\,d)) confirm the presence of NRSS with momentum-dependent spin splitting consistent with the our symmetry analysis.

The layered delafossite structure allows us to systematically study the effects of varying interlayer distance as well as A-site cations, potentially providing an additional degree of control over the spin splitting. We present the results of different interlayer distance in the SI and focus here on the cation intercalation. The crystal structures and bands for cells containing intercalated Li and Na, are shown in Figs. \ref{fig:F4}\,(b,\,e) and (c,\,f), respectively for the $135\degree$, and $109\degree$ structures, and in Fig.~S10\,(b,\,c) in the SI for the $71\degree$ twin boundary. Intercalation of the Li and Na ions causes charge transfer to the adjacent layers, resulting in an increasing of the Co magnetic moments, consistent with previous reports \cite{PhysRevB.77.075119}. However, despite the increased magnetic moments, the magnitude of the NRSS remains relatively similar in both the empty and intercalated structures. This is likely due to the concurrent increase in interlayer distance caused by intercalation, which we found reduces the NRSS as shown in SI Fig. S7, S8 and S9.

While the band structures in Fig.~\ref{fig:F2} and \ref{fig:F4} demonstrate clear NRSS at the DFT level, they rely on artificial periodic boundary conditions that assume a perfectly repeating domain-wall superlattice. In real materials, however, domain walls break translational symmetry, and the concept of a Brillouin zone is no longer strictly applicable. It is therefore essential to assess whether the spin splitting remains robust when moving beyond the idealized periodic crystal limit to more realistic domain sizes. In addition, the density of domain walls is expected to influence the magnitude of the NRSS and its associated spin current. To address these questions, we turn to a transport formalism that does not rely on crystal momentum: using a tight-binding model and the Landauer approach~\cite{Datta1997} implemented in the \textsc{kwant} package~\cite{Groth2014} (details in SI Sec.~1.3), we calculate the spin-resolved conductance as a function of domain size and domain-wall density.

\begin{figure}[t]
\centering
\includegraphics[width=0.95\linewidth]{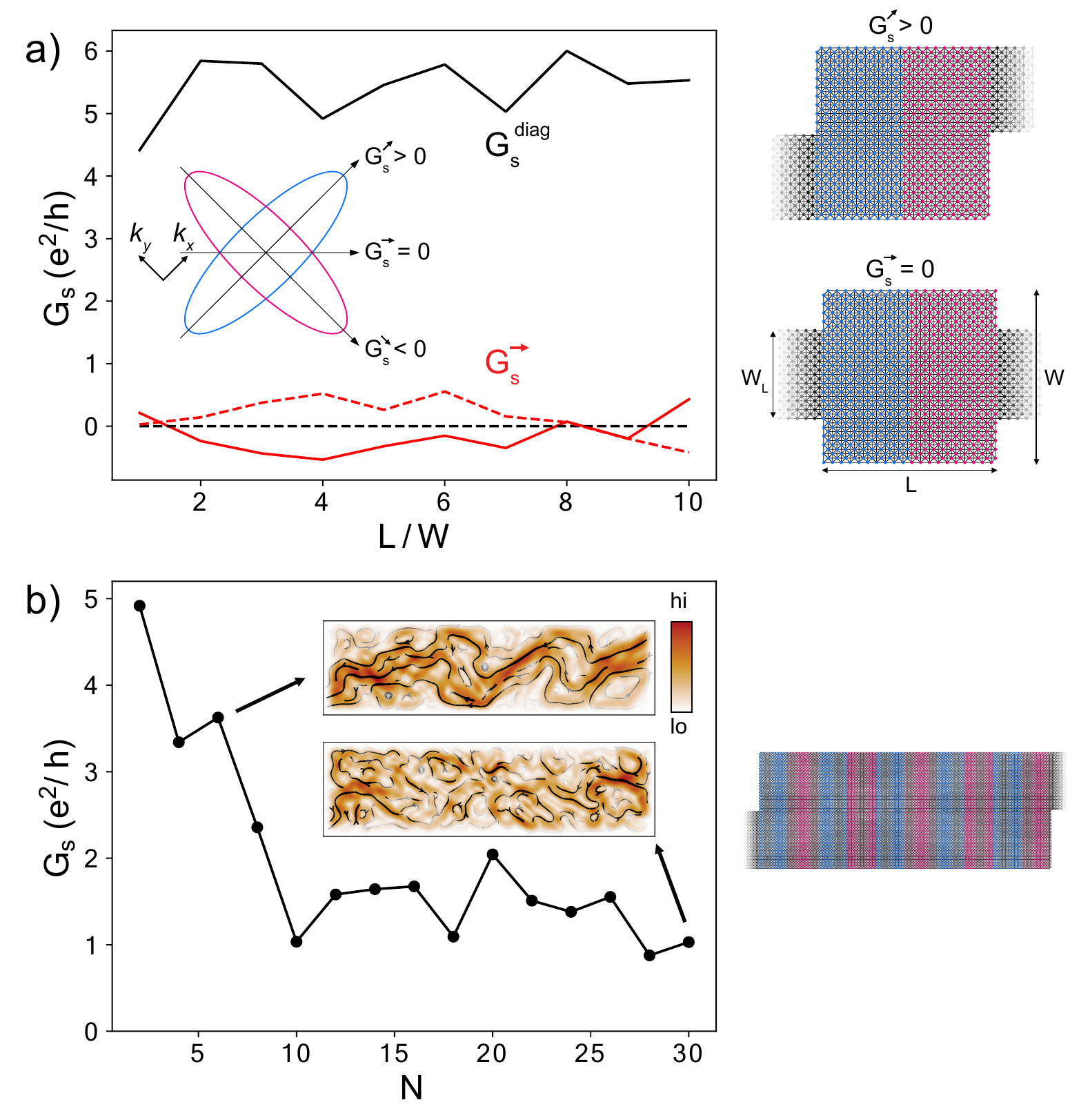}
\caption{(a) $G_s$ as a function of device length for horizontal transport ($G_s^\rightarrow$) and diagonal transport ($G_s^\text{diag} = (G_s^\nearrow - G_s^\searrow)/2$) with (solid) and without (dashed) orbital ordering. Inset shows an schematics of the spin transport anisotropy due to the presence of nonrelativistic spin splitting, where arrows indicate the transport direction. The sketches on the right show the transport setup where half of the device sites have a positive sign of magnetization and orbital ordering (red) and the other half (blue) have a negative sign of magnetization and orbital ordering. Black regions denote the semi-infinite leads. The value and sign of the spin conductance $G_s$ depends on transport direction, dictated by the position of the leads. The device width $W$ is taken as 50 unit cells, the width of the leads as $W_L = W/2$, and the Fermi level $\mu=1.5$. (b) Diagonal spin conductance at a fixed device length as a function of the number of domain walls, $N$. The right schematic shows a device with $N=10$ domains, and the insets show the vector plots of the spin current for $N=6$ and $N=30$. The device size is $L = 200$ and $W = 50$ units cells, and $\mu=1.5$.}
\label{fig:F5}
\end{figure}

In conventional altermagnets, a finite longitudinal spin conductance $G_s$ arises along directions parallel to the momentum-space NRSS axis~\cite{Smejkal2022B}. The sign of $G_s$ reverses under the sublattice-exchanging symmetry operation (rotation or mirror) and vanishes along the nodal-plane directions, as schematically shown in the inset of Fig.~\ref{fig:F5}\,(a). To test this behavior in our domain-wall systems, we compute $G_s$ in the device geometries of Fig.~\ref{fig:F5}\,(a), where the orientation of the leads defines the transport direction. The model comprises two regions with opposite magnetic and orbital order, analogous to the twin-boundary configurations discussed above, and therefore captures the essential symmetry conditions for domain-wall-induced NR

The main panel in Fig.~\ref{fig:F5}\,(a) shows our calculated spin conductance $G_s$ as a function of device length. We evaluate two transport geometries: the horizontal direction ($G_s^{\rightarrow}$) and the diagonal anisotropy defined as $G_s^{\text{diag}} = (G_s^{\nearrow} - G_s^{\searrow})/2$. The results reveal the expected anisotropic signature of NRSS: horizontal transport yields nearly vanishing spin conductance, whereas diagonal transport produces a clear nonzero $G_s^{\text{diag}}$. To confirm that this anisotropy originates from the twin-boundary–induced NRSS, we repeated the calculations with the crystal-field term in the tight-binding model switched off, thereby suppressing the orbital differentiation responsible for NRSS (dashed lines in Fig.~\ref{fig:F5}\,(a)). In this case, both $G_s^{\rightarrow}$ and $G_s^{\text{diag}}$ approach zero, verifying that the nonzero spin conductance is a direct consequence of the domain-wall–induced spin splitting. Moreover, the near-constant dependence of $G_s$ on device length demonstrates that the magnitude of the spin current is insensitive to domain size: even a single twin boundary separating oppositely magnetized domains is sufficient to generate the characteristic anisotropic spin response.

We next explore how the spin conductance evolves with increasing domain-wall density. To do this we fix the total device length and vary the number of alternating magnetic domains $N$, as illustrated in Fig.~\ref{fig:F5}\,(b). The main plot in Fig.~\ref{fig:F5}\,(b) displays $G_s^{\text{diag}}$ as a function of $N$, showing a gradual decay of spin conductance as the number of domain walls increases. To gain microscopic insight into this behavior, we map the real-space spin current across the device (Fig.~\ref{fig:F5}\,(b), insets). For small $N$, when the conductance is high, the current flows smoothly along nearly diagonal trajectories connecting adjacent domains. For large $N$, however, the current paths become irregular and diffuse , indicating that domain walls act as scattering centers that randomize the direction of electron propagation~\cite{Zutic2004}. Because the anisotropic spin conductance relies on electrons coherently traversing diagonally across domain walls, such scattering diminishes the net $G_s^{\text{diag}}$. This behavior contrasts with bulk altermagnets, where anisotropic transport arises from a uniform sublattice symmetry acting throughout the crystal. In the present system, by contrast, the anisotropy originates locally at the twin boundaries, so the cumulative effect of electrons crossing the $N-1$ interfaces dictates the overall spin conductance.

Our results establish a general mechanism by which NRSS an arise in compensated magnetic systems that are not intrinsically altermagnets. The essential ingredient is the change in crystal-field environment across structural domain walls: when neighboring magnetic domains are crystallographically rotated relative to one another, the boundary enforces a sublattice--exchanging operation equivalent to the rotation–-$\mathcal{T}$ or mirror-–$\mathcal{T}$ symmetry that defines conventional altermagnets. In this way, the domains reproduce locally the symmetry conditions required for NRSS, even though the bulk crystal itself remains $\mathcal{PT}$- or $\tau\mathcal{T}$-symmetric.

This mechanism unifies two previously distinct routes to spin splitting. In conventional altermagnets, NRSS originates from intrinsic crystallographic symmetry; at conventional surfaces or interfaces, spin textures arise from the \emph{breaking} of those symmetries. Twin boundaries represent a third regime -- \emph{symmetry reconstruction} -- in which the relative orientation of adjacent domains reintroduces the necessary sublattice relations through a spatially varying but well-defined structural motif. The result is a momentum-dependent spin texture conceptually similar to that of conventional `bulk' altermagnets, but realized entirely through domain geometry. At the microscopic level, the effect relies on the same unifying principle as conventional NRSS: the coexistence of compensated magnetic order and orbital symmetry differentiation. Across a domain boundary, variations in crystal-field orientation or bonding topology produce alternating orbital occupations on opposite magnetic sublattices. This interplay between lattice geometry, orbital character, and magnetic compensation generates the observed spin splitting without the need for relativistic spin–-orbit coupling.

Twin boundaries and ferroic domain walls are known to strongly modify local electronic and ferroic responses in a broad range of materials~\cite{sluka_enhanced_2012a,sluka_freeelectron_2013a,faraji_electrical_2017,Seidel2009,eggestad,PhysRevMaterials.2.114405, Moore_et_al:2022,Meier2022}. Our findings extend this understanding by showing that when the structural motif changes across domains, the resulting domain configurations can provide precisely the sublattice--exchanging symmetry operation required for NRSS, even in bulk-symmetric crystals. Crucially, the NRSS is not confined to an atomically narrow interface but is a symmetry property of the full domain architecture: the boundary sets the relative crystallographic operations between domains, and the ensuing orbital symmetry differentiation generates a momentum-dependent spin texture with two orthogonal nodal planes. This symmetry--dictated texture produces a distinct transport anisotropy, vanishing $G_s$ along the nodal directions and finite spin conductance along diagonal trajectories across domains, offering clear experimental signatures.

In multiferroic \ce{BiCoO3}, the coexistence of ferroelastic and ferroelectric order provides additional tunability. The polar distortion across the $90^{\circ}$ head-to-tail walls subtly differentiates the Co environments on either side, yielding small opposite local moments that cancel macroscopically. Because these moments are coupled to the ferroelectric polarization, an applied electric field could in principle reverse the local spin asymmetry while preserving the overall NRSS symmetry, enabling electric-field control of the spin response in multiferroic systems.

The delafossite \ce{CoO2} system, by contrast, highlights how structural degrees of freedom, such as interlayer spacing or cation intercalation, can be used to tune the magnitude of the spin splitting. The sensitivity of NRSS to these geometric parameters implies that external pressure, strain, or chemical modification can modulate the effect continuously. This tunability connects domain-wall-induced NRSS to the broader class of ferroelastic and shape-memory materials, where twin formation and detwinning can be reversibly controlled~\cite{bhattacharya_microstructure_2003}. In such systems, a mechanically detwinned state would suppress NRSS by restoring $\mathcal{PT}$ or $\tau\mathcal{T}$ symmetry, whereas a re-twinned state would reinstate it, suggesting mechanically reconfigurable spin functionality (see SI Fig.~S11).

Unlike spin-–orbit--based mechanisms such as Rashba or Dresselhaus effects, our approach achieves spin polarization purely through exchange interactions and domain symmetry, establishing a general, SOC--free route to spin functionality in compensated magnets. Because twin boundaries are ubiquitous and experimentally accessible, the predicted NRSS could be probed using spin angle--resolved photoemission spectroscopy (sARPES) or spin angle--resolved reflection electron spectroscopy (sARRES) to map the momentum-dependent spin texture~\cite{Dale2024}, or by anisotropic magnetotransport to detect the characteristic spin-current anisotropy \cite{Gonzalez2021, GonzalezBetancourt2024, Noh2025}.

In summary, we have demonstrated that twin boundaries can induce and control NRSS in materials where such effects are otherwise forbidden. Using \ce{BiCoO3} and \ce{CoO2} as representative examples, we find a $d$-wave-type NRSS with two symmetry-protected nodal planes, robust anisotropic spin transport independent of domain size, and a systematic reduction with increasing domain-wall density. These findings establish structural-domain engineering as a general symmetry framework for realizing spin--polarized electronic states in the absence of spin–-orbit coupling, opening new opportunities for electrically and mechanically tunable spintronic devices.

\section{Acknowledgements}
This work was supported by the U.S. Department of Energy, Office of Science, Office of Basic Energy Sciences, Materials Sciences and Engineering Division under Contract No.\ DE-AC02-05CH11231 within the Theory of Materials program. Computational resources were provided by the National Energy Research Scientific Computing Center and the Molecular Foundry, DOE Office of Science User Facilities supported by the Office of Science, U.S.\ Department of Energy under Contract No.\ DE-AC02-05CH11231 and Sigma2 – the National Infrastructure for High-Performance Computing and Data Storage in Norway through project NN9264K. The work performed at the Molecular Foundry was supported by the Office of Science, Office of Basic Energy Sciences, of the U.S.\ Department of Energy under the same contract. Support for this project was provided by the Research Council of Norway (Project no. 302506).

\clearpage

\bibliographystyle{unsrt}
\bibliography{references}

\end{document}

%%%%%%%%%%%%%%%%%%%%%%%%%%%%%%%%%%%%%%%%%%%%%%%%%%%%%%%%%%%%%%%%%%%%

\title{{\Large Supplementary Information}\\[-0.4 em]Twin-boundary-induced nonrelativistic spin splitting}

\author{Kristoffer Eggestad}
\affiliation{Department of Materials Science and Engineering,\\[-0.8 em] NTNU - Norwegian University of Science and Technology, NO-7491 Trondheim, Norway}

\author{Marc Vila}
\affiliation{Materials Sciences Division, Lawrence Berkeley National Lab,\\[-0.8 em] 1 Cyclotron Road, Berkeley, CA 94720, USA}
\affiliation{Molecular Foundry, Lawrence Berkeley National Lab,\\[-0.8 em] 1 Cyclotron Road, Berkeley, CA 94720, USA}

\author{Sverre M. Selbach}
\email{selbach@ntnu.no}
\affiliation{Department of Materials Science and Engineering,\\[-0.8 em] NTNU - Norwegian University of Science and Technology, NO-7491 Trondheim, Norway}

\author{Sinéad M. Griffin}
\email{sgriffinh@lbl.gov}
\affiliation{Materials Sciences Division, Lawrence Berkeley National Lab,\\[-0.8 em] 1 Cyclotron Road, Berkeley, CA 94720, USA}
\affiliation{Molecular Foundry, Lawrence Berkeley National Lab,\\[-0.8 em] 1 Cyclotron Road, Berkeley, CA 94720, USA}

\sloppy

\maketitle

\beginsupplement

\section{Computational Methods and Additional calculations}
First-principles calculations were performed using the \texttt{VASP} code\cite{vasp1,vasp2,vasp3}. Interactions between core and valence electrons (Bi: (5d$^{10}$, 6s$^{2}$, 6p$^3$), Co: (3s$^{2}$, 3p$^6$, 3d$^{8}$, 4s$^{1}$), O: (2s$^2$, 2p$^4$)) were treated using the projector-augmented wave method (PAW)\cite{PhysRevB.50.17953,paw}. PBEsol \cite{PBEsol} + U were used for all calculations with a U value of 3.3 eV, benchmarked against magnetic moments on Co ions calculated using the HSE06 functional\cite{HSE06}, as the HSE06 functional is known to give accurate descriptions of magnetism and electronic structure of transition metal oxides. A plane-wave energy cutoff of 600 eV and gamma centered grids with \textit{k}-spacing of less than $0.25$ Å$^{-1}$ were used for all calculations. 

\subsection{Twin Boundary Supercells}

\begin{figure}[ht]
    \centering
    \includegraphics[width=0.8\linewidth]{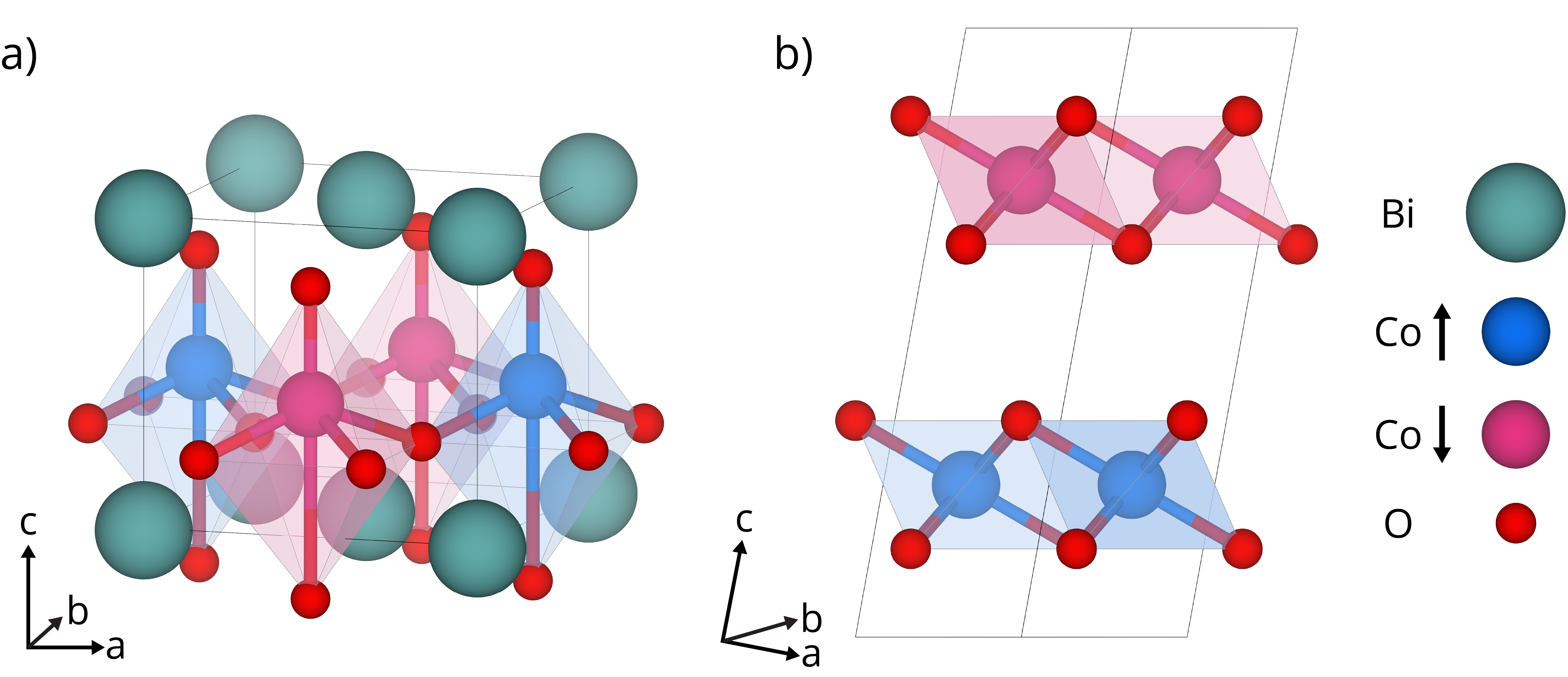}
    \caption{Primitive structures of \ce{BiCoO3} (a) and \ce{CoO2} (b) optimized using the PBEsol functional using the magnetic configurations indicated by the coloured atoms and octahedra.}
    \label{fig:structures}
\end{figure}

Supercells containing twin boundaries were created using the DFT optimized lattice parameters of the primitive structures, optimized with a force criterion of 0.01 eV/Å$^2$. For \ce{BiCoO3}, we used C-type antiferromagnetic setting, while in the case of \ce{CoO2}, ferromagnetic order within each layer and antiferromagnetic coupling between layers were used. The DFT-optimized lattice constants are presented in SI Table \ref{tab:lattice_PBEsol} and the optimized structures with magnetic order in SI Figure \ref{fig:structures}. 

\clearpage

\setlength{\tabcolsep}{0.7em}
\begin{table}[ht]
    \centering
    \caption{Lattice parameters for \ce{BiCoO3}, \ce{CoO2}, \ce{LiCoO2} and \ce{NaCoO2} optimized using the PBEsol functional.}
    \begin{tabular}{ccccccc}
         & a (Å) & b (Å) & c (Å) & $\alpha$ ($\degree$) & $\beta$ ($\degree$) & $\gamma$ ($\degree$) \\
         \midrule
        \ce{BiCoO3} & 5.22 & 5.22 & 4.69 & 90 & 90 & 90 \\
        \ce{CoO2} & 2.79 & 2.79 & 9.10 & 98.8 & 98.8 & 60 \\
        \ce{LiCoO2} & 2.90 & 2.90 & 9.64 & 98.7 & 98.7 & 60 \\
        \ce{NaCoO2} & 2.99 & 2.99 & 10.7 & 98.1 & 98.1 & 60 \\
    \end{tabular}
    
    \label{tab:lattice_PBEsol}
\end{table}

To estimate twin boundary energies, supercells containing 16 formula units were used. All of which were modelled with ferromagnetic order to exclude contributions from magnetic DWs and isolate the intrinsic energy of the twin boundaries.

\subsection{Electronic Structure}
\begin{figure}[ht]
    \centering
    \includegraphics[width=0.90\linewidth]{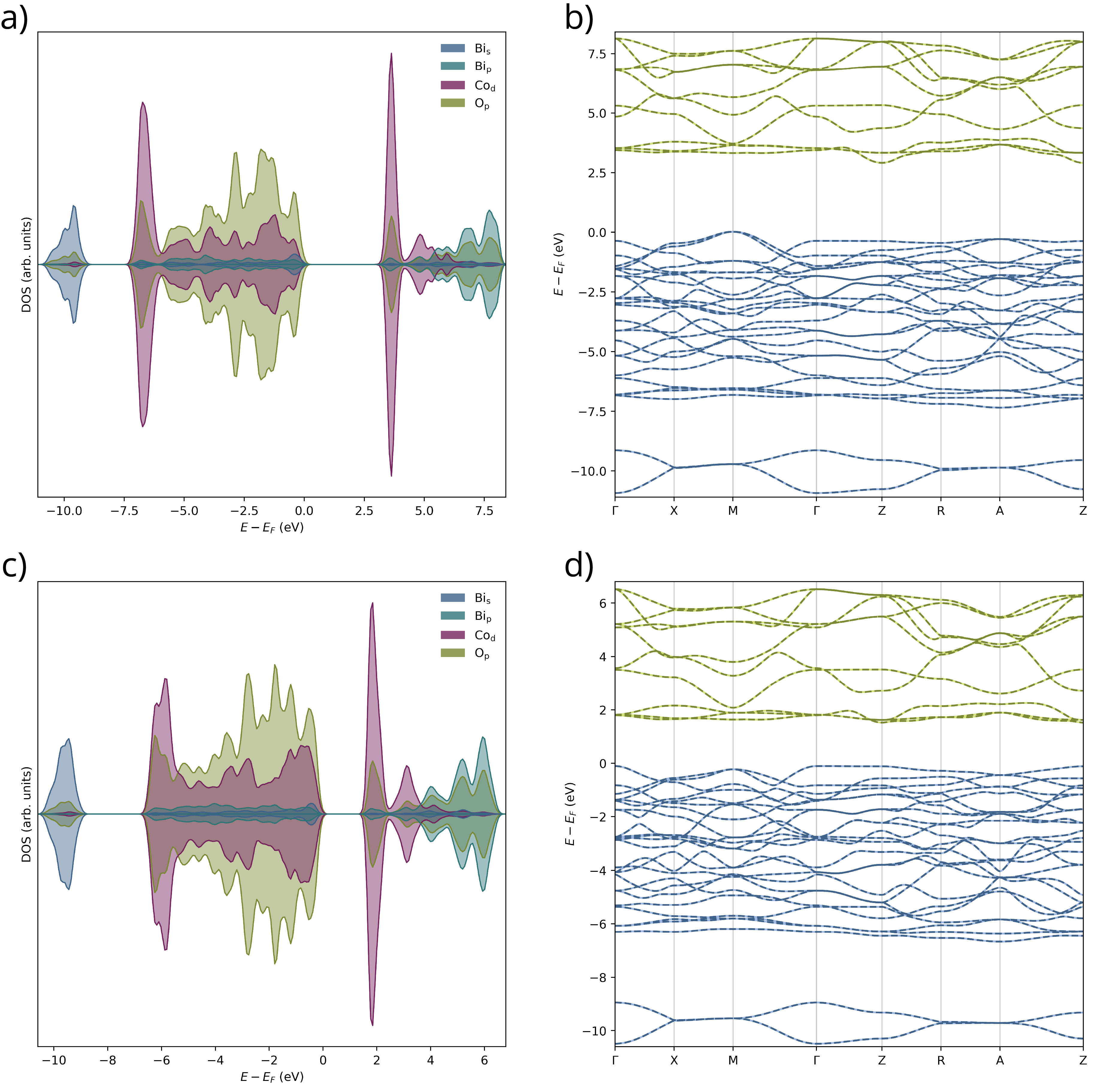}
    \caption{Electronic density of states (a and c) and band structures (b and d) for \ce{BiCoO3} calculated using the HSE06 (a and b) and PBEsol (c and d) functional.}
    \label{fig:electronic_bulk_BiCoO3}
\end{figure}

\begin{figure}[ht]
    \centering
    \includegraphics[width=0.90\linewidth]{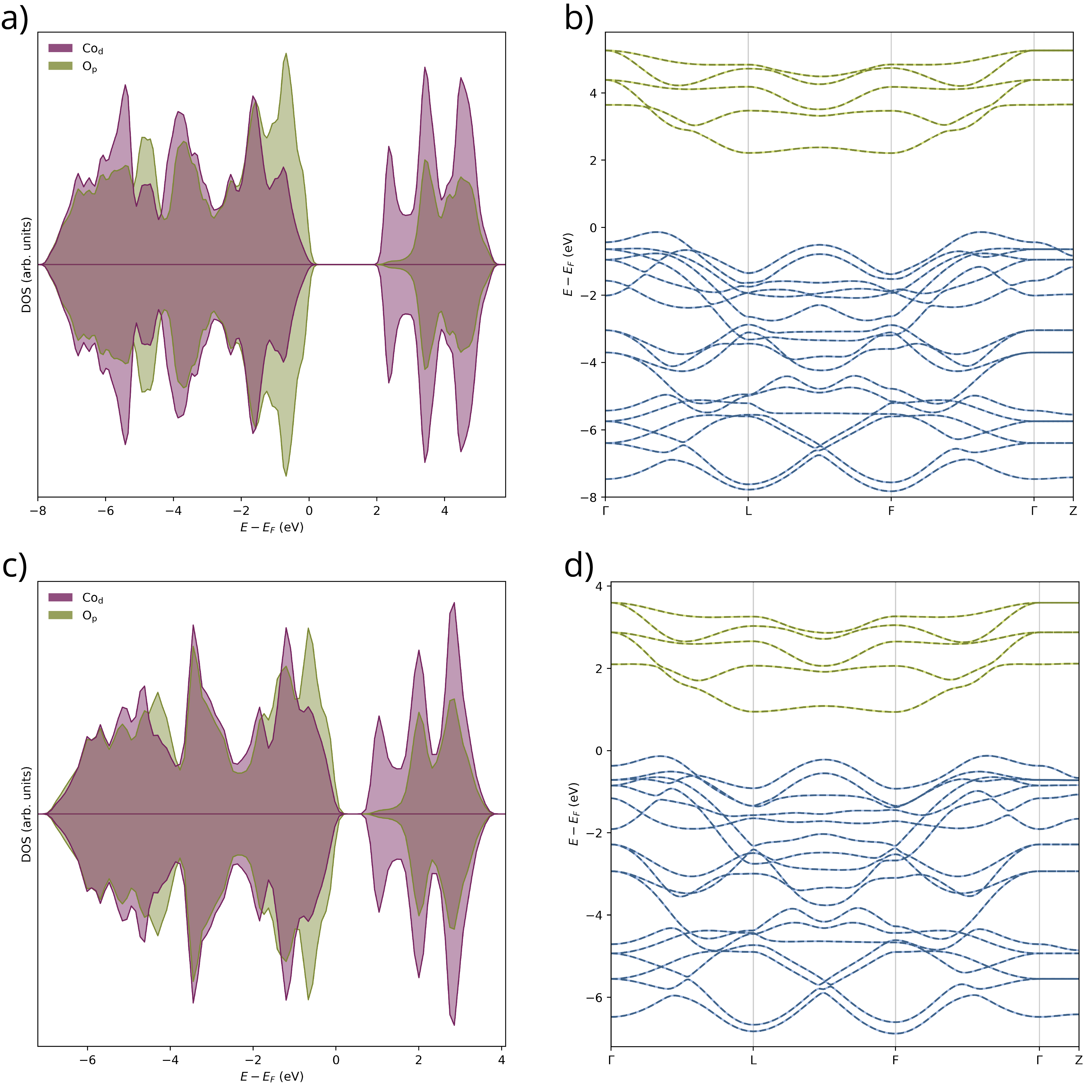}
    \caption{Electronic density of states (a and c) and band structures (b and d) for \ce{CoO2} calculated using the HSE06 (a and b) and PBEsol (c and d) functional.}
    \label{fig:electronic_bulk_CoO2}
\end{figure}

\begin{figure}[ht]
    \centering
    \includegraphics[width=0.90\linewidth]{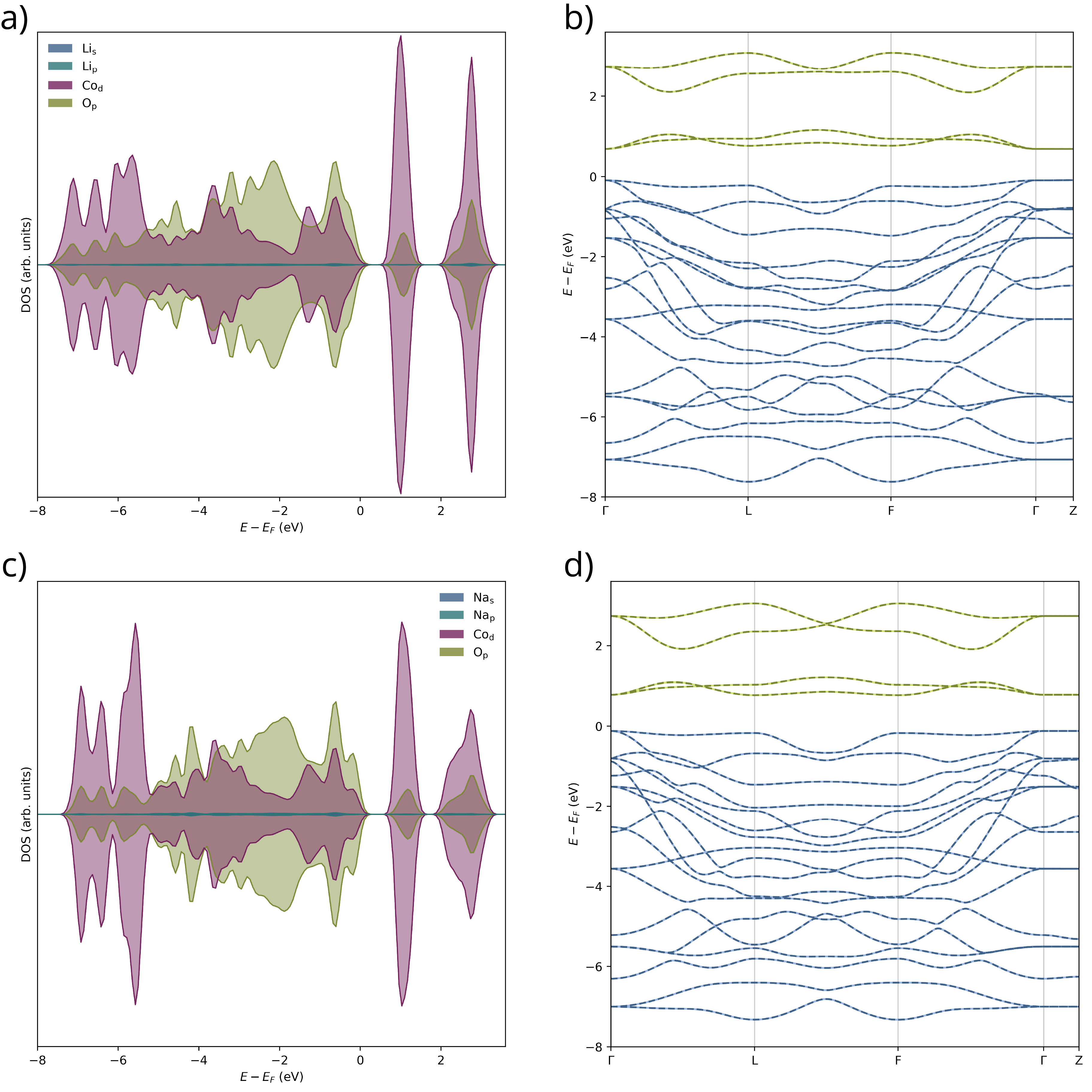}
    \caption{Electronic density of states (a and c) and band structures (b and d) for \ce{LiCoO2} (a and b) and \ce{NaCoO2} (c and d) calculated using the PBEsol functional.}
    \label{fig:electronic_bulk_CoO2_LiNa}
\end{figure}

All electronic structure calculations were performed using an electronic convergence criterion of $10^{-8}$ eV. The $k$-point paths used for band structure calculations are specified in the respective figures. Calculated electronic structure of pristine \ce{BiCoO3} and \ce{CoO2} using both HSE06 and PBEsol+U, displayed in SI Figure \ref{fig:electronic_bulk_BiCoO3} and \ref{fig:electronic_bulk_CoO2}, shows that the functional gives relatively similar results. Apart from the underestimated band gaps, PBEsol predicts the same bonding environment as HSE06. The electronic structure of Li and Na intercalated \ce{CoO2} are presented in SI Figure~\ref{fig:electronic_bulk_CoO2_LiNa}.

\clearpage

\begin{figure}[ht]
\centering
\includegraphics[width=0.9\linewidth]{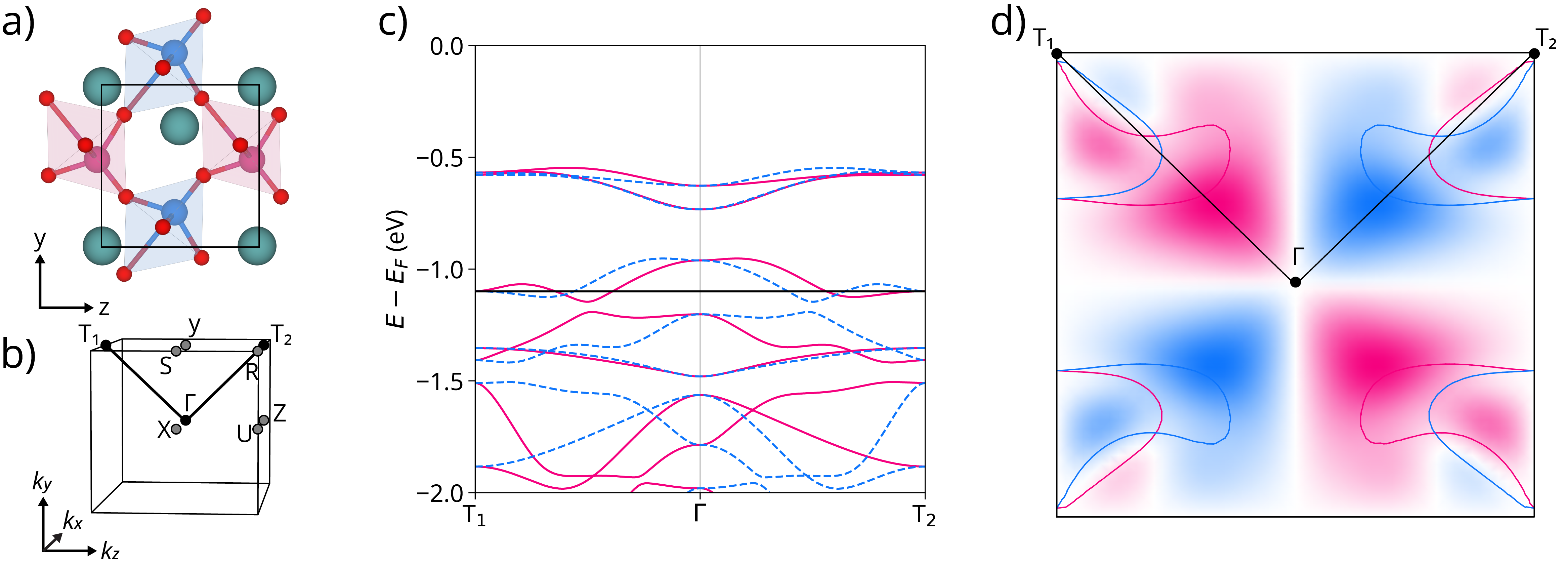}
\caption{(a) Minimal engineered domain wall structure of \ce{BiCoO3}. (b) Corresponding Brillouin zone, where the high-symmetry path ($\text{T}_1 \rightarrow \Gamma \rightarrow \text{T}_2$) used in the band structure is indicated. (c) Electronic band structure along the Brillouin zone path shown in (b). The solid black line marks the energy at $-1.1$ eV below the Fermi level. Constant energy cut in the \textit{ky}-\textit{kz} plane at $-1.1$ eV below the Fermi level. The solid coloured lines represent intersections of spin-up and spin-down bands with the energy cut (black line in (c)), and the contour plot shows the broadened bands as calculated using Equation \ref{eq:broad}.}
\label{fig:splitting_BiCoO3}
\end{figure}

Spin-resolved constant energy surface plots were computed by identifying the intersections of spin-up and spin-down bands at a specific energy, $\mu$, with reference to the Fermi level. Numerically, this quantity is calculated as the product of the spin polarization and a delta function at energy $\mu$ approximated by a Gaussian function:

\begin{equation}
    \begin{aligned}
        s(k,\mu) &= \sum_n \delta(E_n(k) - \mu) \langle n, k | s | n, k \rangle \\
        & \approx \sum_n e^{-\frac{(E_n(k) - \mu)^2}{2\sigma^2}} \langle n, k | s | n, k \rangle,
    \end{aligned}
\label{eq:broad}
\end{equation}

in units of $\hbar/2$. Here, $E_n$ is the energy of band $n$, $k$ is a specific \textit{k}-point and $\sigma$ is the smearing parameter which is set to $\sigma = 0.02$ eV for Figure~\ref{main-fig:F2}~(d) and $\sigma = 0.08$ for SI Figure~\ref{fig:splitting_BiCoO3}~(d). This approach highlights the symmetry of the spin splitting in $k$-space, making it easier to identify the location of nodal surfaces.

\begin{figure}[ht]
    \centering
    \includegraphics[width=0.95\linewidth]{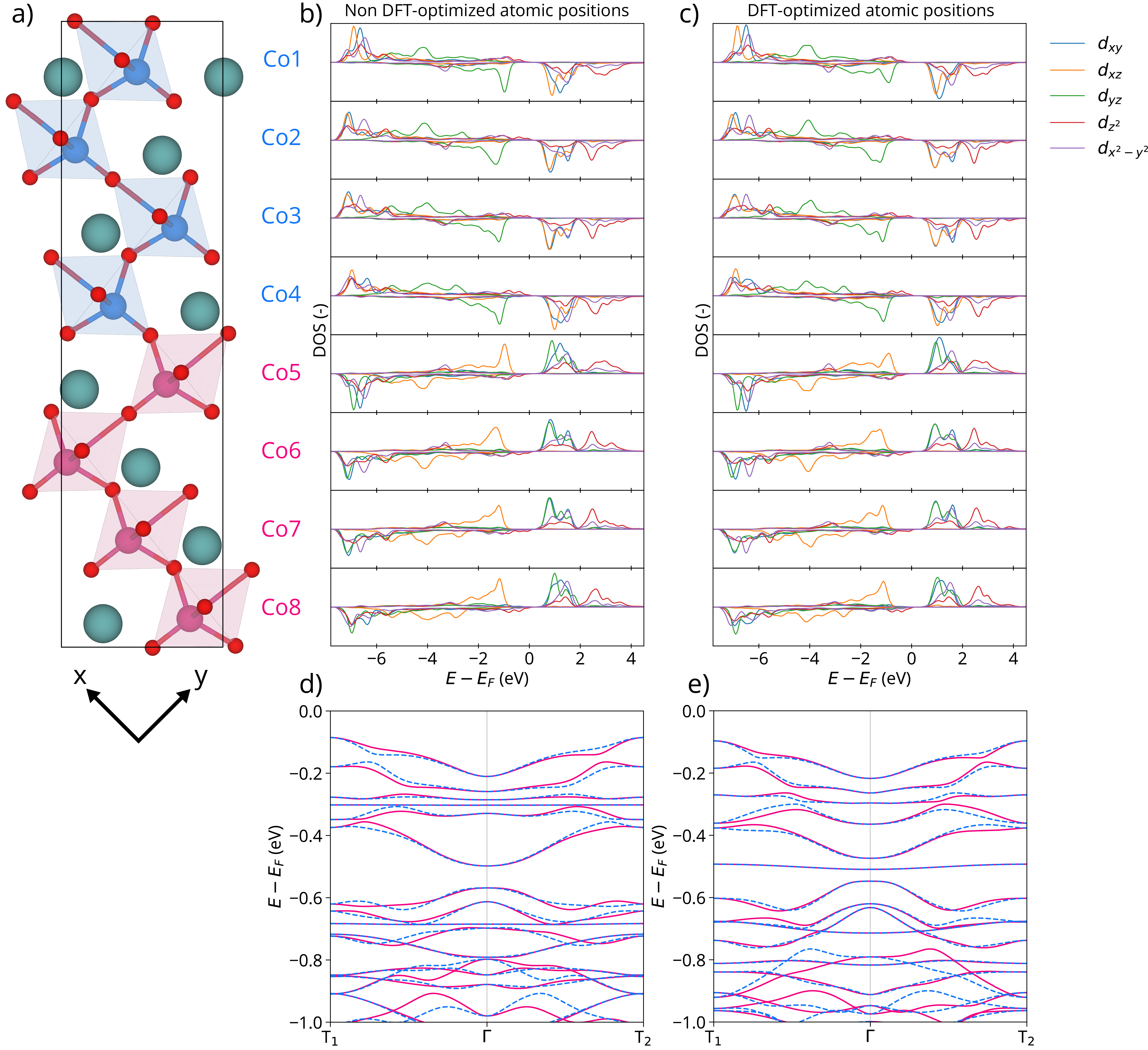}
    \caption{(b) and (c) show d-orbital projected density of states for Co atoms in the non-optimized and DFT optimized structure displayed in (a). This layer resolved DOS, highlighting the orbital ordering in each domain. (d) and (e) show the calculated band structure for the same structures following the path shown in Figure \ref{main-fig:F2} in the main text.}
    \label{fig:orbital_ordering_8}
\end{figure}

The d-orbital projected electronic density of states (DOS) presented in Figures~\ref{main-fig:F3} and \ref{fig:orbital_ordering_8} were obtained using the \texttt{lobster} software package \cite{doi:10.1021/jp202489s,https://doi.org/10.1002/jcc.23424,https://doi.org/10.1002/jcc.26353}, with input files generated with \texttt{lobsterpy}\cite{Naik2024}. Before the \texttt{lobster} calculations, the unit cells were reoriented as depicted in the corresponding figures.

Within the central regions of each domain (Co number 2, 3, 6 and 7), the two half-filled $t_{2g}$ orbitals remain nearly spin-degenerate, consistent with the behavior observed in the ground state structure (Figure~\ref{main-fig:F3}~(a)). The slight deviation from perfect spin degeneracy is attributed to the chosen axis orientation, which does not perfectly align with the local structure in either domain. Approaching the domain wall, these orbitals become increasingly non-degenerate due to the alternating crystalline structure as in Figure~\ref{main-fig:F3}~(b).
\\\\
Additionally, SI Figures \ref{fig:orbital_ordering_8}~(c) and (e) present the layer-resolved DOS and band structure, respectively, with atomic positions optimized using PBEsol. The results are nearly identical to those shown previously, indicating that structural optimization of the domain wall has a negligible effect on the observed spin splitting.

\subsubsection{Effect of Interlayer Spacing on NRSS}

\begin{figure}[ht]
    \centering
    \includegraphics[width=0.95\linewidth]{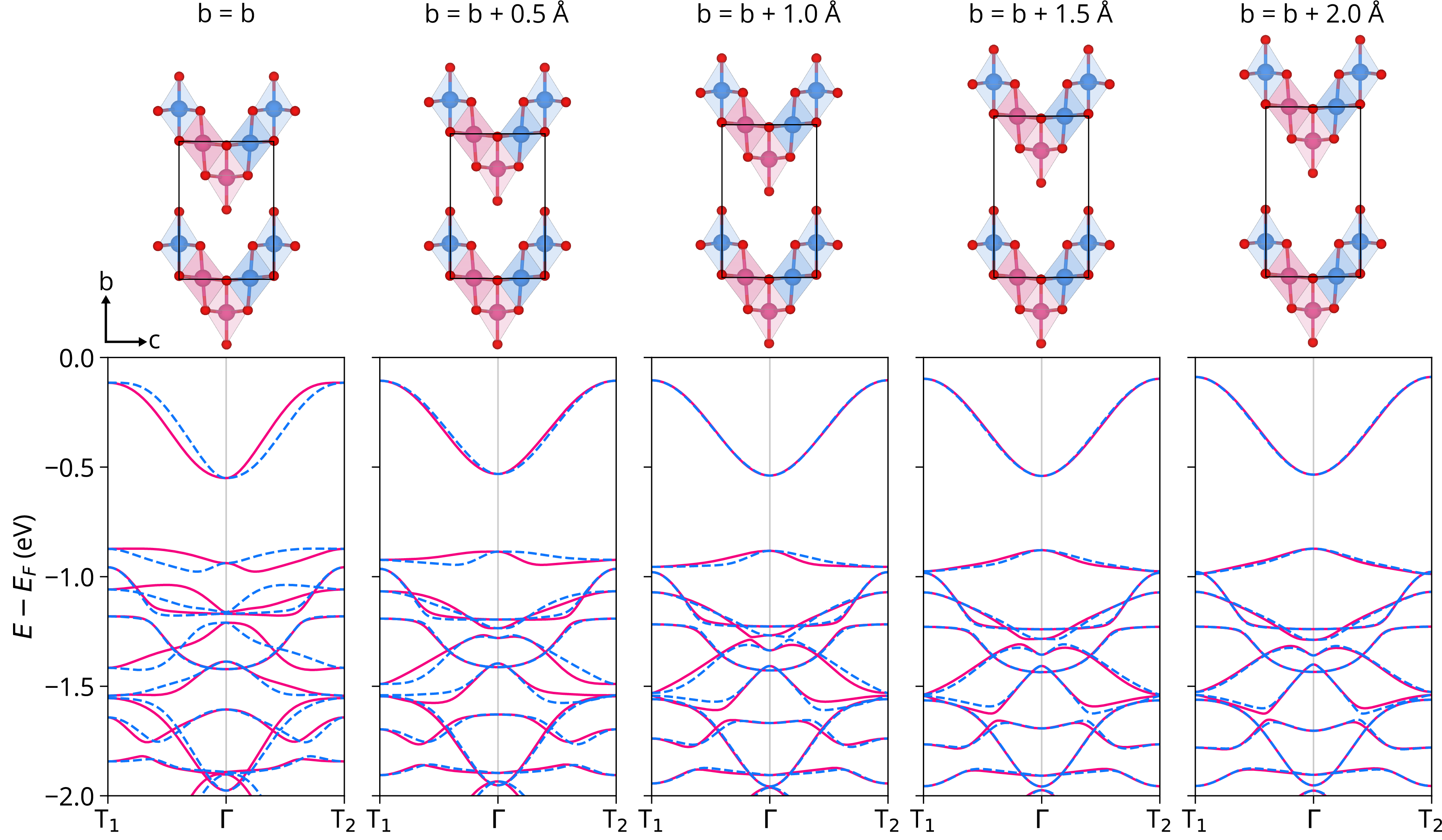}
    \caption{Band structures showing the evolution of the momentum-dependent spin-splitting as a function of interlayer spacing for $71\degree$ twin boundaries in \ce{CoO2}.}
    \label{fig:interlayer_71}
\end{figure}

\begin{figure}[ht]
    \centering
    \includegraphics[width=0.95\linewidth]{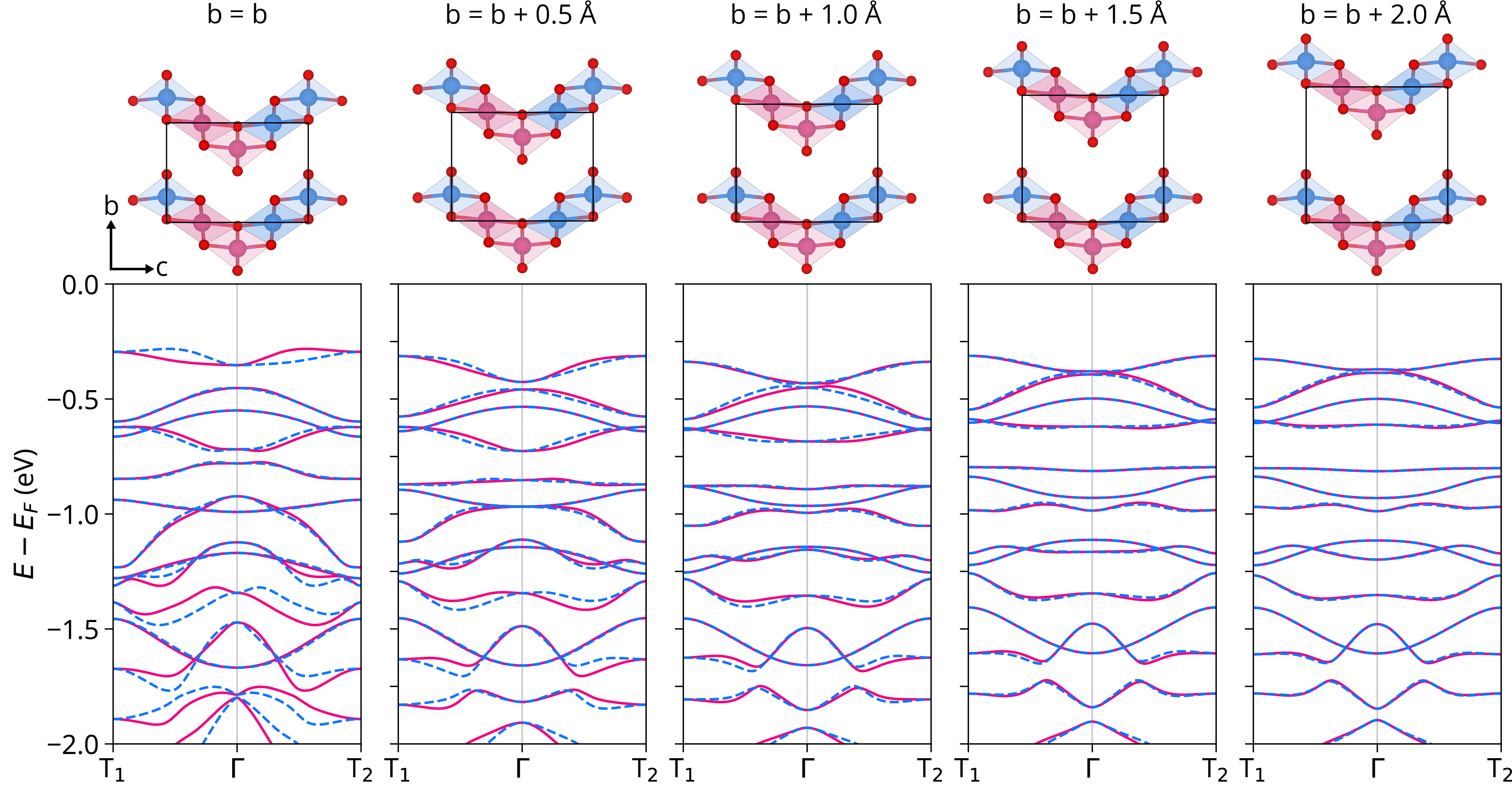}
    \caption{Band structures showing the evolution of the momentum-dependent spin-splitting as a function of interlayer spacing for $109\degree$ twin boundaries in \ce{CoO2}.}
    \label{fig:interlayer_109}
\end{figure}

\begin{figure}[ht]
    \centering
    \includegraphics[width=0.95\linewidth]{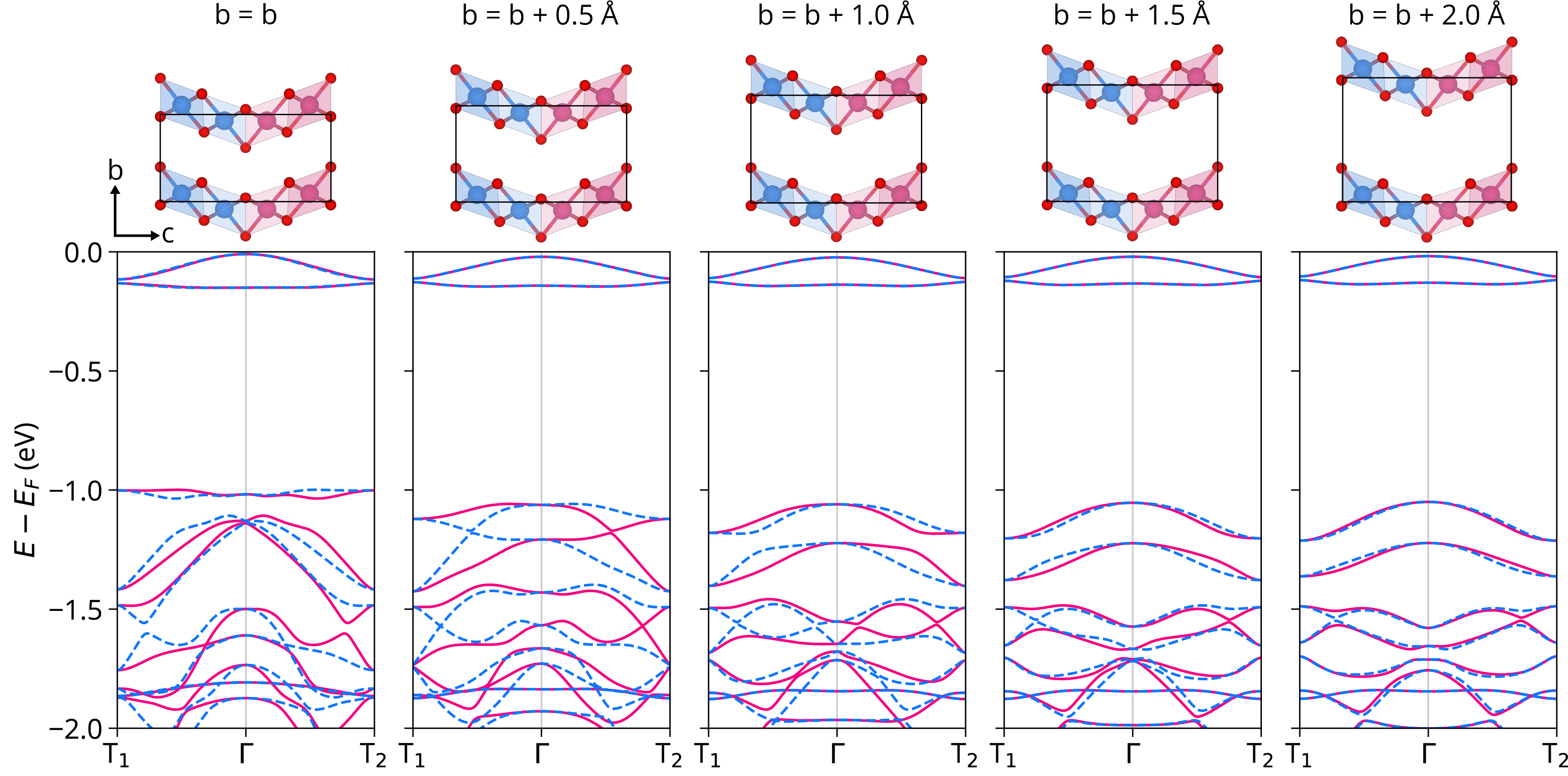}
    \caption{Band structures showing the evolution of the momentum-dependent spin-splitting as a function of interlayer spacing for $135\degree$ twin boundaries in \ce{CoO2}.}
    \label{fig:interlayer_135}
\end{figure}

SI Figure \ref{fig:interlayer_71}, \ref{fig:interlayer_109} and \ref{fig:interlayer_135} show how the band structures of the $71\degree$, $109\degree$ and $135\degree$ twin boundaries in \ce{CoO2} evolve as a function of increasing interlayer distance. 
\\\\
Notably, we find that the spin splitting reduces with increasing interlayer distance. As the interlayer distance increases, interactions between layers are suppressed, causing the band dispersion along $k_y$ to become progressively flatter, while bands between $k_y$ and $k_z$ increasingly reflect only the $k_z$ direction. Since $k_y=0$ corresponds to a nodal plane, the spin splitting vanishes as interplanar coupling is lost. A comparable trend has been reported for interchain spacing in monolayers of 1D materials such as \ce{VBr3} and \ce{CrCl3} \cite{zhds-vnyt}. 

\clearpage
\subsubsection{Effect of Intercalation of Ions in Layered Systems on NRSS}

\begin{figure}[ht]
    \centering
    \includegraphics[width=0.9\linewidth]{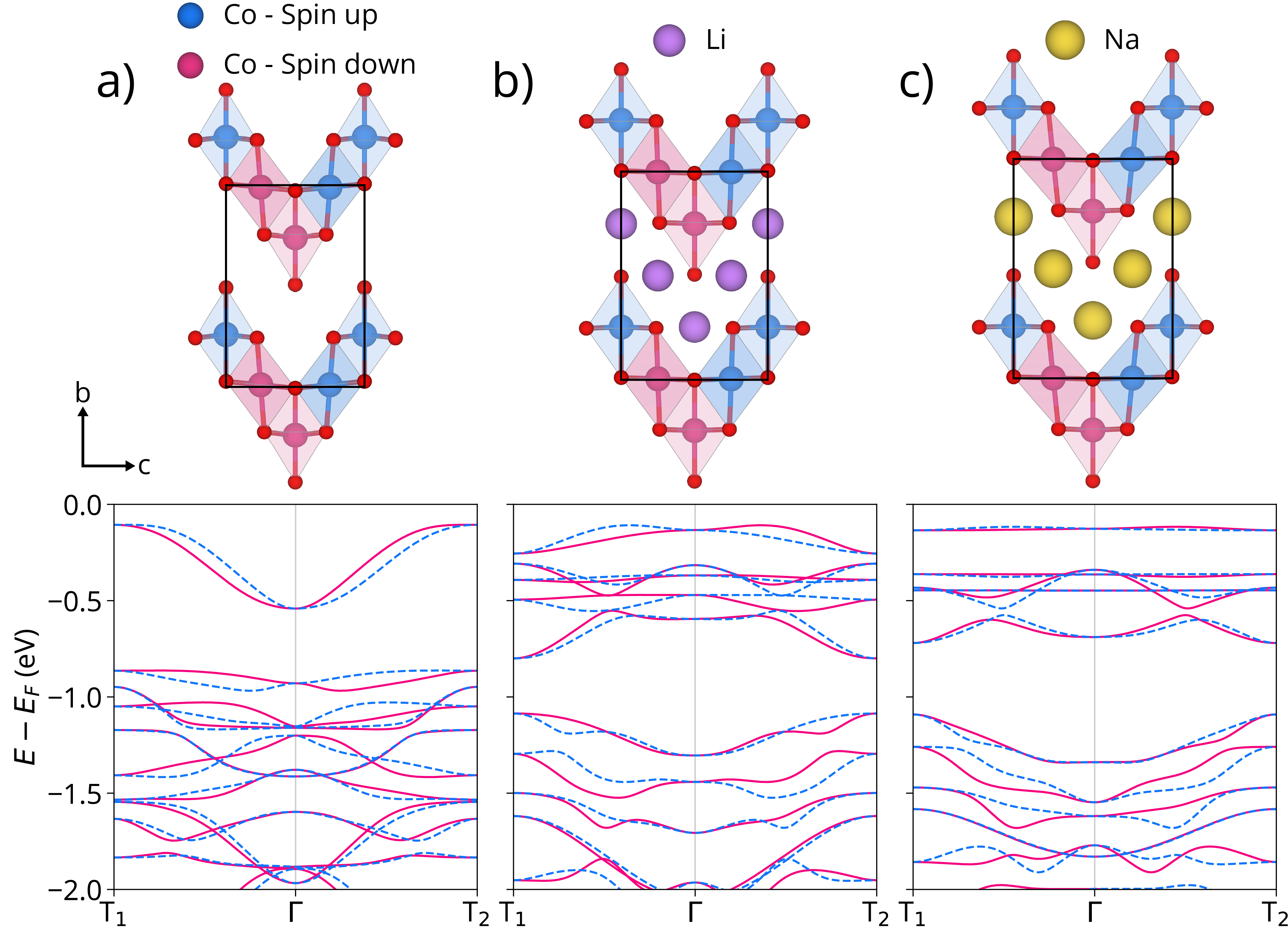}
    \caption{Electronic band structure showing a path with momentum-dependent spin-splitting for $71\degree$ twin boundaries in empty \ce{CoO2} (a) and \ce{CoO2} intercalated with Li (b) and Na (c).}
    \label{fig:bs_71}
\end{figure}

Figure \ref{main-fig:F4} and SI Figure \ref{fig:bs_71} show how the band structures of the $71\degree$, $109\degree$ and $135\degree$ twin boundaries in \ce{CoO2} change with the introduction of Li and Na ions. Both structures \ce{LiCoO2} and \ce{NaCoO2} show a robust NRSS present in all the calculated twin boundaries. The intercalation of either Li or Na effectively reduces Co ions and results in significantly larger magnetic moments.

\clearpage

\subsection{Transport calculations} \label{section_transport}

Spin-resolved conductances and spatially-resolved spin currents were calculated within the Landauer-B\"{u}ttiker formalism using the Kwant package \cite{Groth2014}. We implemented a tight-binding model based on the d-wave altermagnet model from Ref. \cite{Vila2025}, which, in addition to the momentum-dependent spin splitting, also captures orbital ordering via crystal field effects. It is a two-dimensional square lattice, with two sublattices, and two orbitals ($d_{xz}, d_{yz}$) and spin per site:

\begin{equation}
    \begin{aligned}
        \mathcal{H} &= \sum_{\langle i,j \rangle, \tau, \tau^\prime} c_{i\tau}^\dagger [t_{ij}]_{\tau\tau^\prime} c_{j\tau^\prime} + m_{ex} \sum_{i,s,s^\prime} c_{is}^\dagger [\textbf{m}_i \cdot \textbf{s}]_{ss^\prime} c_{is^\prime} \\ 
        & + \sum_{i,\tau, \tau^\prime} c_{i\tau}^\dagger [\Delta_i]_{\tau\tau^\prime} c_{i\tau^\prime}  + H.C.
    \end{aligned}
\label{eq_H0} 
\end{equation}

Here, $c^\dagger$ and $c$ are the creation and annihilation operators, respectively, and sites are labelled by the index $i$, spin $s$ and orbital $\tau$. The first term denotes the hopping between nearest-neighbor sites with strength $t$. It is derived based on the Slater-Koster method \cite{Slater1954} with parameters $V_\pi$ and $V_\delta$ characterizing the nature of the chemical bonds. The second term describes a collinear, compensated magnetic exchange interaction in the bipartite lattice, $m_{ex}$, where spins $\textbf{s}$ of electrons couple to localized magnetic moments $\textbf{m}_i$. We take $\textbf{m} || \hat{\textbf{z}}$ for concreteness. The third term describes the crystal field $\Delta$ that splits the energy of the orbitals. For tetragonal distorted octahedra in a square lattice, as in the case of \ce{BiCoO3}, $\Delta_i = + \Delta$ for octahedron aligned along the $x$ axis, and $\Delta_i = - \Delta$ for octahedron aligned along the $y$ axis (see coordinate frame in Figure \ref{main-fig:F3} (b)) \cite{Vila2025}. The parameters for the calculations are chosen as $V_\pi = -1$, $V_\delta=0.3$, $m_{ex} = 0.7$, $\Delta = 0.4$ in units of $|V_\pi|$.
\\\\
To simulate the multiferroic domain wall structure (Figure 4 in the main text), we model half of the device with a positive sign of magnetization and crystal field, and another half with a negative sign of these parameters. The Hamiltonian of the leads have the same hopping parameters as that of the main device region, but without crystal field and magnetization. This is chosen so that any effect related to crystal fields or magnetism comes solely from the scattering region. 
\\\\
The spin conductance is defined as the difference of charge conductance between up and down spins:
\begin{equation}
G_s = \frac{e^2}{h} (G_\uparrow - G_\downarrow),
\end{equation}
where $e$ is the electron charge and $h$ the Planck constant.

\subsection{Twin Boundary Energies}
\label{sec:twin_energy}

The calculated energies of $90\degree$ head-to-tail, head-to-head/tail-to-tail, and $180\degree$ ferroelectric DWs in \ce{BiCoO3} are summarised in Table \ref{tab:dw_energy}, alongside the energies of the various twin boundaries in \ce{CoO2}. All calculations were performed in a ferromagnetic configuration in order to isolate the energetic contribution of the twin boundary itself.
\\\\
Consistent with results for isostructural \ce{PbTiO3}\cite{PhysRevB.65.104111}, the $90\degree$ head-to-tail ferroelectric and ferroelastic DW in \ce{BiCoO3} exhibits significantly lower energy than the head-to-head/tail-to-tail $90\degree$ and $180\degree$ DWs. This indicates that the $90\degree$ head-to-tail configuration is the most energetically favourable and thus the most likely to form in \ce{BiCoO3}. However, the energy of this DW remains substantially larger than reported values for other oxides \cite{eggestad,PhysRevLett.110.267601,PhysRevB.65.104111}, which is expected due to the pronounced tetragonal distortion and concomitant strain field. For \ce{CoO2}, the $109\degree$ twin boundary is found to have the lowest energy, in agreement with experimental observations \cite{JIANG2020105364,exp_twin}.

\begin{table}[ht]
\centering
\caption{Twin boundary energies calculated using the PBEsol functional for supercells containing 16 formula units. "$180\degree$ 1" and "$180\degree$ 2" denote DWs oriented parallel to one of the $\langle 100 \rangle$ or $\langle 110 \rangle$ planes, respectively, in the primitive perovskite unit cell.}
\begin{tabular}{ccc}
Compound & Type & Energy (mJ/m$^2$) \\
\toprule
\multirow{ 3}{*}{\ce{BiCoO3}} & $90\degree$ HT & 174 \\
& $90\degree$ HH/TT & 1021 \\
& $180\degree$ 1 & 1290 \\
& $180\degree$ 2 & 372 \\
\midrule
\multirow{ 3}{*}{\ce{CoO2}} & $71\degree$ & 970 \\
& $109\degree$ & 548 \\
& $135\degree$ & 1456 \\
\bottomrule
\end{tabular}

\label{tab:dw_energy}
\end{table}

\clearpage

\section{Shape Memory Altermagnet}

\begin{figure}[bht]
    \centering
    \includegraphics[width=0.9\linewidth]{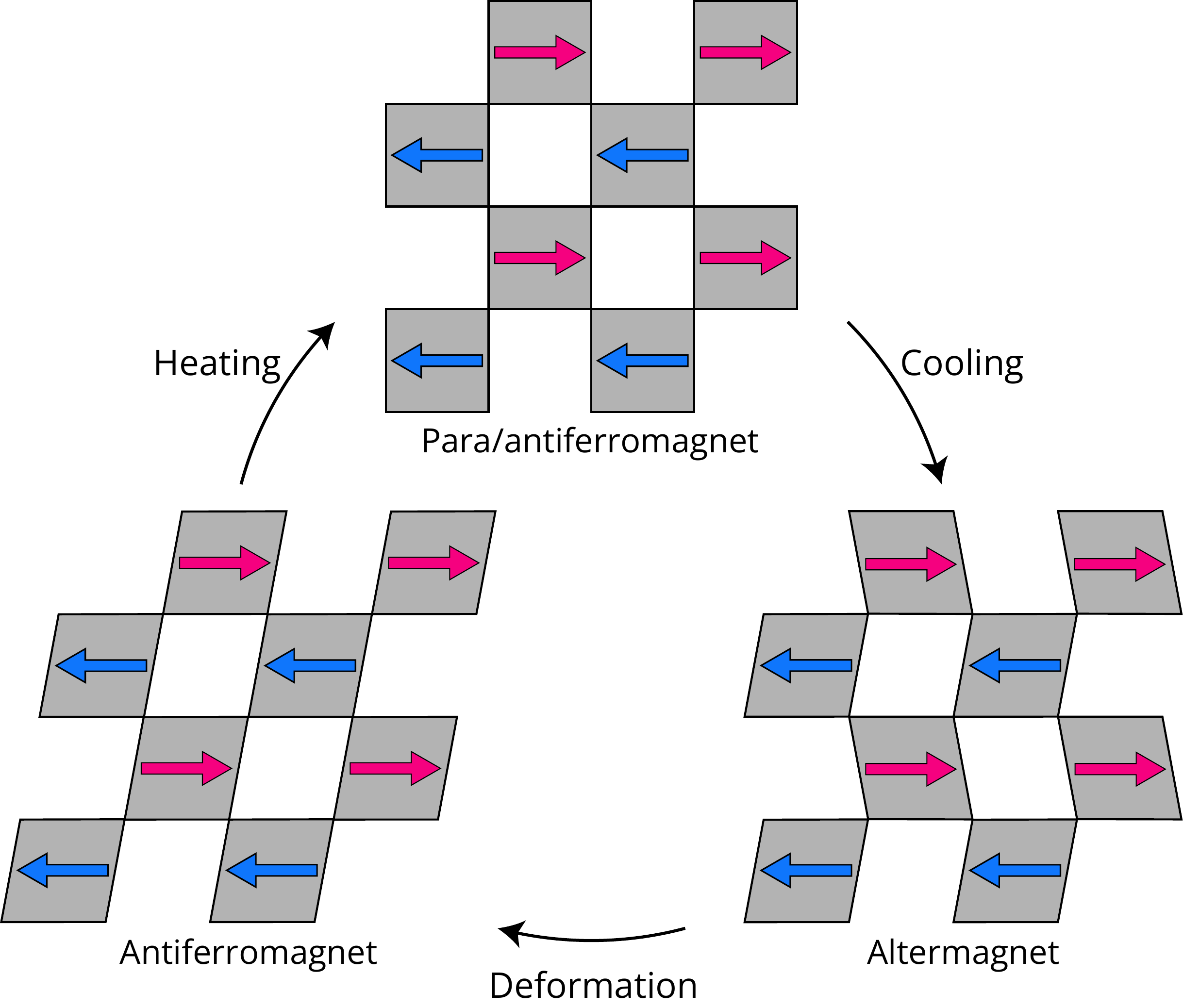}
    \caption{Sketch of the working principle of a shape memory altermagnet.}
    \label{fig:shape_memory_magnet}
\end{figure}

The sketch shown in SI Figure~\ref{fig:shape_memory_magnet} illustrates the working principle of a shape memory altermagnet. If a ferromagnetic shape memory material is engineered such that its twinned state exhibits momentum dependent NRSS, as demonstrated in this work, the material’s altermagnetic properties can be switched on and off. Specifically, the twinned (altermagnetic) state can be converted to a detwinned (non-altermagnetic) state through mechanical deformation, and the altermagnetism can be restored by another heat treatment cycle. 

\clearpage

\begin{figure}[ht]
    \centering
    \includegraphics[width=0.9\linewidth]{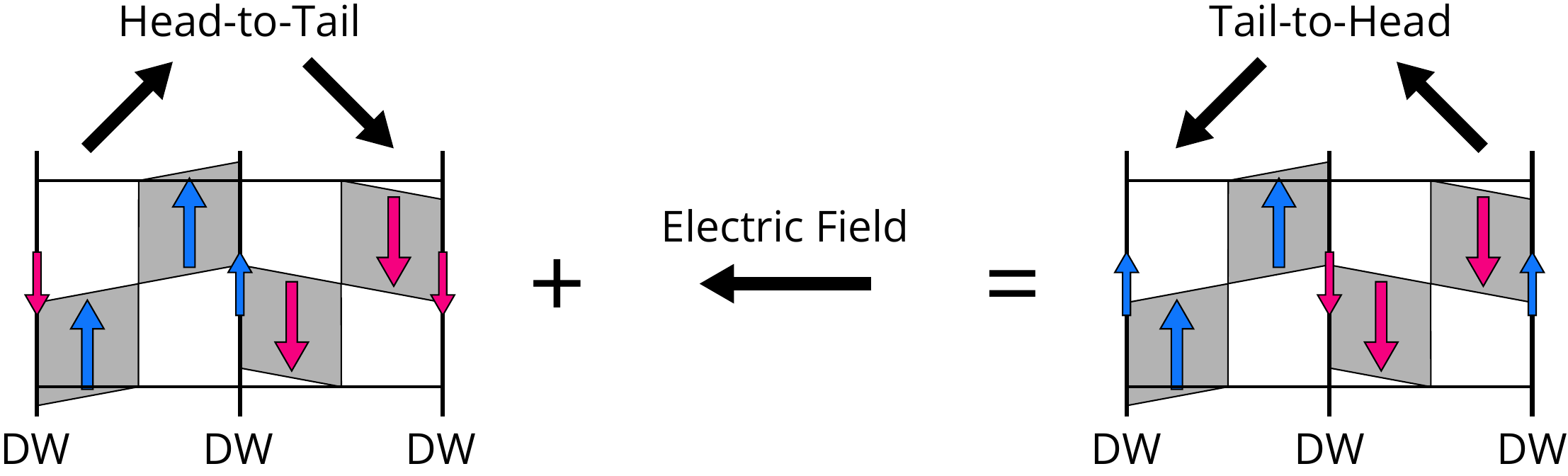}
    \caption{Sketch showing how the magnetic moment of a head-to-tail ferroelectric domain wall can be switched by an external electric field.}
    \label{fig:switching}
\end{figure}

Figure \ref{fig:switching} shows how small residual magnetic moments, that arise at head-to-tail domain walls, potentially can be controlled by applying an external electric field to switch the polarization direction.

\footnotesize

\bibliographystyle{unsrt}
\bibliography{references}